\documentclass[onecolumn]{article}
\newtheorem{definition}{Definition}

\usepackage{amssymb}
\usepackage{amsmath}
\usepackage{graphicx}
\usepackage[tight,footnotesize]{subfigure}
\usepackage{algorithm}
\usepackage{algorithmic}

\begin{document}

\title{Trust beyond reputation: A computational trust model based on stereotypes\footnote{Note: Liu Xin and Krzysztof Rzadca did this work at Nanyang Technological University, Singapore.}}

\author{Xin Liu\\ \'{E}cole Polytechnique F\'{e}d\'{e}rale de Lausanne (EPFL)\\ x.liu@epfl.ch \and
Anwitaman Datta\\ Nanyang Technological University\\ anwitaman@ntu.edu.sg \and
Krzysztof Rzadca\\ University of Warsaw\\ krzadca@mimuw.edu.pl}

\date{}

\maketitle

\begin{abstract}

Models of computational trust support users in taking decisions. They are commonly used to guide users' judgements in online auction sites; or to determine quality of contributions in Web~2.0 sites.
However, most existing systems require historical information about the past behavior of the specific agent being judged. In contrast, in real life, to anticipate and to predict a stranger's actions in absence of the knowledge of such behavioral history, we often use our ``instinct''---essentially stereotypes developed from our past interactions with other ``similar'' persons.
In this paper, we propose StereoTrust, a computational trust model inspired by stereotypes as used in real-life. A stereotype contains certain features of agents and an expected outcome of the transaction. When facing a stranger, an agent derives its trust by aggregating stereotypes matching the stranger's profile. Since stereotypes are formed locally, recommendations stem from the trustor's own personal experiences and perspective. Historical behavioral information, when available, can be used to refine the analysis.
According to our experiments using Epinions.com dataset, StereoTrust compares favorably with existing trust models that use different kinds of information and more complete historical information.

\end{abstract}


\section{Introduction}
\label{sec:introduction}

Trust is an important abstraction used in diverse scenarios including e-commerce~\cite{corbitt2003trust, kim2011more}, distributed and peer-to-peer systems~\cite{powertrust,adaptive} and computational clouds~\cite{kaufman2009data, cachin2009trusting}. 
Since these systems are open and large, participants (agents) of the system are often required to interact with others with whom there are few or no shared past interactions. To assess the risk of such interactions and to determine whether an unknown agent is worthy of engagement, these systems usually offer some trust-management mechanisms to facilitate decision support.

If a trustor has sufficient direct experience with an agent, the agent's future performance can be reliably predicted~\cite{Mui02acomputational}. However, in large-scale environments, the amount of available direct experience is often insufficient or even non-existent. In such circumstances, prediction is often based on trustor's ``indirect experience'' --- opinions obtained from other agents that determine the target agent's reputation~\cite{eigen03,Xiong04peertrust,distributedtrust97}.
Simple aggregations, like a seller's ranking on eBay, rely on access to global information, like the history of the agent's behavior. Alternatively, \emph{transitive trust models} (or web of trust models)~\cite{distributedtrust97,transitive03} build chains of trust relationships between the trustor and the target agent. The basic idea being that if $A$ trusts $B$ and $B$ trusts $C$, then $A$ can derive its trust in $C$ using $B$'s referral on $C$ and $A$'s trust in $B$. In a distributed system with many agents and many interactions, constructing such chains requires substantial computational and communication effort. Additionally, if the referral trust is inaccurate, the transitive trust may be erroneous.
Furthermore, trust may not be transitive between different contexts: $A$ may trust $B$ in a certain context (e.g., serving good food), but not necessarily to recommend other agents (a ``meta''-context, or the referral context)~\cite{Xiong04peertrust,Swamynathan05decouplingservice}.

All the above mentioned approaches aggregate \emph{the same kind of information} --- agents' impressions about past transactions, i.e. behavioral history. However, many systems provide a vast context for each transaction, including its type, category, or participants' profile. We were curious to see how accurately one could predict trust using such contextual information. The primary objective of this work was thus not to outperform existing trust models using the same information. In contrast, we explore a complementing alternative that can enhance the existing models --- or to replace them, when the vast information they require is not available (for instance, as a bootstrapping mechanism).

The contribution of this work is a trust model that estimates target agents' trust using stereotypes learned by the assessor from its own interactions with other agents having similar profile.
Our work is partly inspired by~\cite{Tadelis,Tirole96} that study the relation between the reputation of a company and its employees:
The company's reputation can be modeled as an aggregate of employees' reputations and it can suggest a prior estimate of an employee's reputation.
In StereoTrust, agents form stereotypes by aggregating information from the context of the transaction, or the profiles of their interaction partners. Example stereotypes are ``programmers working for Google are more skilled than the average'' or ``people living in good neighborhoods are richer than the others''. To build stereotypes, an agent has to group other agents (``programmers working for Google'' or ``people living in good neighborhoods''). These groups do not have to be mutually exclusive. Then, when facing a new agent, in order to estimate the new agent's trust, the trustor uses groups to which the new agent belongs to.

In case an agent does not have enough past experience to form stereotypes, we propose to construct a stereotypes sharing overlay network (SSON), which allows the newcomers to get stereotypes from established agents. In Section~\ref{sec:share} we discuss how a newcomer can then use the stereotypes without completely trusting the agents who produced them.

StereoTrust assumes that the trustor is able to determine the profile of the target agent, and that the extracted information is accurate enough to distinguish between honest and dishonest agents.\footnote{Example environments where StereoTrust may be applied include (1) identifying unknown malicious sellers in online auction sites~\cite{stereotrustonlineauction}, (2) selecting reliable peers to store replicated data in Peer-to-Peer storage systems~\cite{stereotrustStorage}, (3) filtering out junk emails or prioritizing coming emails by studying features of the old emails,
to name a few.} The profile of an agent is a construct that represents all the information a trustor can gather. An underlying assumption is that interaction with an agent about which no information is available is rare. For instance, when interacting with people, the profile may be constructed from the agent's Facebook profile. Similarly, a profile of a company may be constructed from its record in Companies Registry or Yellow Pages. Generally, an agent's profile is hard to forge as it is maintained by a third party (e.g., historical transaction information of an eBay seller). In case the profile is manipulated (e.g., in a decentralized environment), the trustor may resort to distributed security protection mechanisms or even to human operators to extract the correct information. Note that StereoTrust will likely determine as useless the parts of the profile which can be easily forged and do not have sufficient discriminating power.
While having its own weaknesses and limitations, we think that StereoTrust is interesting both academically and in practice. Academically, its novelty lies in emulating a human behavior of modeling trust by stereotypes for the first guess about a stranger.
In practice, as, first, StereoTrust may be applicable in scenarios in which information used by traditional trust models is not available, noisy, inaccurate or tampered. Second, StereoTrust provides personalized recommendations based on a trustor's own experience (in contrast to the ``average'' experience used in reputation-based approaches). Third, if the global information used by a standard trust models is available, it can be seamlessly integrated to our model in order to enhance the prediction (thus, the term ``stereo'' in the model's name is a double entendre). The stereotypes are improved by dividing the original group into ``honest'' and ``dishonest'' subgroups based on available data about past behavior of agents.
Our experiments show that such a dichotomy based refinement, called d-StereoTrust, significantly improves the accuracy.

Since StereoTrust defines groups in a generic way, it may be used in very different kinds of applications; even within the same application, different agents may have their own personal, locally defined groups. Also, the notion of trust itself can be easily adapted to different contexts. In this paper, we use the widely-adopted definition of trust as the trustor's subjective probability that a target agent will perform a particular action~\cite{canwetrusttrust}.


As an example application, consider judging the quality of product reviews from a web site such as Epinions.com. In such a community, users write reviews for products, structured into different categories (e.g., books, cars, music). These reviews are later ranked by other users. Normally, each reviewer has some categories in which she is an ``expert'' (like jazz albums for a jazz fan). The reviewer is more likely to provide high quality reviews for products in these familiar categories. Of course, users may also write reviews for products from other categories, but their quality might be inferior, because of, e.g., insufficient background knowledge. ``Mastery'' can be correlated between categories. For instance, audiophiles (people interested in high-quality music reproduction) usually know how to appreciate music; thus, if they decide to review a jazz album, the review is more likely to be in-depth. The correlation might also be negative, as one may not expect an insightful review of a jazz album from, e.g., a game boy reviewer. When facing an unrated review of a jazz album by an unknown contributor, we can use the information on the contributor's past categories (game-boy fan or an audiophile?) and our stereotypes (``noisy'' gamers vs. insightful audiophiles) to estimate the quality of the review. We use Epinions.com data to evaluate StereoTrust (Section~\ref{sec:evaluation}): we demonstrate that taking into account reviewers' interests provides a good estimation of the quality of the review.

Consider a very different kind of application --- a peer-to-peer storage system. If a peer wants to store a new block of data, it needs to choose a suitable replication partner. The suitability of a partner peer depends on the likelihood that the peer will be available when the data needs to be retrieved (which may depend on its geographic location, time-zone difference, etc.); the response time to access the data (which may depend on agreements and infrastructure between internet service providers); and many other factors. Existing systems usually use a multi-criteria optimization model, and thus need substantial knowledge about the specific peer in question --- for instance, its online availability pattern, end-to-end latency and bandwidth. Applying StereoTrust can provide an alternative systems design, where a peer in, say Tokyo, can think --- ``my past experiences tell me that peers in Beijing and Hong Kong have more common online time with me compared with peers in London and New York. Likewise, peers in New York and Hong Kong with a specific IP prefix provide reliable and fast connections, while the others don't.'' Based on such information, if the peer has to choose between a partner in Hong Kong or in London, the peer can make the first guess that a peer in Hong Kong is likely to be its best bet,  without needing to know the history of the specific peers in question. A ``mixed'' data placement strategy that uses available historical information and stereotypes is expected to result in even better performance. It is worth noting that we are not claiming that it necessarily provides the best possible system design, but merely that it opens the opportunity for alternative designs. We have in fact devised such StereoTrust guided data placement strategy for P2P backup systems, resulting in several desirable properties compared to other placement strategies \cite{stereotrustStorage}.


The core of our contribution is the basic StereoTrust model (Section~\ref{basicmodel}) and d-StereoTrust, the dichotomy based enhanced model using historical information (Section~\ref{sec:enhencedmodel}).
We also discuss how to select features to form stereotypes (Section~\ref{preliminary}). Feature preprocessing can help to form discriminating stereotypes, and thus to improve the accuracy of the trust assessment. When the trustor does not have sufficient past transactions to form stereotypes, we propose to construct an overlay network to share stereotypes among agents to help inexperienced agents bootstrap the system (Section~\ref{sec:share}).
An experimental evaluation using a real-world (Epinions.com) and a synthetic dataset (Section~\ref{sec:evaluation}) shows that stereotypes provide an adequate ``first guess'' --- and when coupled with some historical data (d-StereoTrust), the resulting trust estimates are more accurate than the estimates provided by the standard trust models.

\section{Related Work}
\label{sec:relatedwork}

Intuitively, past mutual interactions are the most accurate source to predict agent's future behavior \cite{Mui02acomputational}, but such an approach is unsuitable in large-scale distributed systems where an agent commonly has to assess a target agent with whom it has no past interactions.

Instead of using only local experience, many approaches derive the trust from target agent's reputation --- information aggregated from other agents. For instance, \cite{distributedtrust97} and \cite{transitive03} derive trust from paths of trust relationships starting at the trustor (the asking agent), passing through other agents and finishing at the target agent (the transitive trust model). However, the transitive trust model has several drawbacks, such as handling wrong recommendations, efficiently updating trust in a dynamic system, or efficiently establishing a trust path in a large-scale system.

EigenTrust \cite{eigen03} is a reputation system developed for P2P networks. Its design goals are self-policing, anonymity, no profit for newcomers, minimal overhead and robustness to malicious coalitions of peers. EigenTrust also assumes transitivity of trust. The peers perform a distributed calculation to determine the eigenvector of the trust matrix. The main drawback of EigenTrust is that it relies on some pre-trusted peers, which are supposed to be trusted by all peers.
This assumption is not always true in the real world. First, these pre-trusted peers become points of vulnerability for attacks. Second, even if these pre-trusted peers can defend the attacks, there are no mechanisms to guarantee that they will be always trustworthy. Additionally, EigenTrust (and some other reputation systems like \cite{Xiong04peertrust}) is designed based on Distributed Hash Tables (DHTs), thus imposing system design complexity and deployment overheads. In contrast, our proposed model does not rely on any specific network structure for trust management.

\cite{delegation06} proposed a trust system using groups. A group is formed based on a particular interest criterion; group's members must follow a set of rules. The approach assumes that the leader of each group creates the group and controls the membership. The trust is calculated by an aggregative version of EigenTrust, called Eigen Group Trust. In Eigen Group trust, all the transactions rely on the group leaders, who are assumed to be trusted and always available. This approach requires certain special entities (i.e., group leaders) to coordinate the system, thus may suffer from scalability issue, while StereoTrust has no such restrictions.

REGRET~\cite{regret} combines direct experience with social dimension of agents, that also includes so-called system reputation. System reputation is based on previous experience with other agents from the same \emph{institution}. Unlike StereoTrust's stereotypes, REGRET's institutions exist outside the system, thus each agent can be assigned to an institution with certainty. REGRET also assumes that each agent belongs to a single institution.

BLADE \cite{blade} derives trustworthiness of an unknown agent based on the feedback collected from other agents. This model treats agent's properties (i.e., certain aspects of the trustworthiness) and other agents' feedback as random variables. By establishing a correlation between the feedback and the target agent's properties using Bayesian learning approach, the trustor is able to infer the feedback's subjective evaluation function to derive certain property of the unknown target agent.

The related works reviewed so far mainly rely on direct experience and/or indirect experience with the target agent to derive trust. In contrast, StereoTrust uses another kind of information --- stereotypes learned from the trustor's own interactions with \emph{other} agents having similar profiles. So StereoTrust provides a complementing alternative that can enhance the existing models, particularly when the vast information they require is not available.

While using stereotypes for user modeling and decision making was suggested previously  \cite{usermodelingstereotype,reputationsociety}, to the best of our knowledge the conference version of this paper~\cite{stereotrust09} is the first concrete, formal computational trust model that uses stereotypes. Several other papers on similar lines have been published since then, demonstrating the potential of using stereotypes for assessing trust. \cite{bootstrapStereotype} used stereotypes to address the cold-start issue. The trustor constructs stereotypes relying on M5 model tree learning algorithm \cite{learningcontinuousclass}. That is, stereotype construction is treated as a classification problem (i.e., learning association between features and the expected probability of a good outcome), so no explicit groups are built. In contrast, our approach forms and maintains explicit groups and infers the corresponding stereotypes based on aggregated behaviors of the group members. This makes the derived stereotypes easy to be interpreted by the real users.
Similarly to our Stereotype-Sharing Overlay Network, in their work, new trustors can request stereotypes from the experienced ones, and then combine these stereotypes with the target agent's reputation (if any) using subjective logic \cite{subjectivetrust}. Although similar, our approach has several advantages: (1) as a basic trust model, StereoTrust uses the (well adapted) beta distribution to derive stereotypes and trust, thus the resulting decision is easy to be interpreted; (2) The work \cite{bootstrapStereotype} shares agents' local stereotypes heuristically (e.g., no discussion on how stereotype providers are selected) while StereoTrust offers a more sophisticated sharing mechanism by maintaining a dedicated overlay network for exchanging, combining, and updating stereotypes; (3) StereoTrust is applicable in practice as demonstrated in real-world dataset (Epinions.com) based evaluation (see Section \ref{sec:evaluation}) presented in this paper, as well as in other diverse applications and settings in which we have applied stereotype to derive trust \cite{stereotrustStorage,stereotrustonlineauction}.

\cite{stereotypesource} considers the problem of identifying useful features to construct stereotypes. Three feature sources are discussed: (i) From the social network, i.e., relationships between agents are features; (ii) From agents' competence over time. The target agent's accumulated experience in certain tasks can be viewed as features. An example stereotype may be \emph{if an agent performed task T more than 100 times, he is considered experienced (trustworthy)}. (iii) From interactions, e.g., features of both interaction parties. For instance, the trustor with certain features is positively or negatively biased towards the target agent with other features.
This work provided a comprehensive summary of feature sources (for stereotype formation) from social relationships among agents, but the authors did not apply these features to any concrete application scenarios for validation. Such works identifying suitable features for building stereotypes complement StereoTrust's abstract computational trust model leveraging on said stereotypes.

The stereotypes StereoTrust uses can be also regarded as a generalization of various hand-crafted indirect trust or reputation metrics---alternatively, these metrics can be interpreted as complex stereotypes (in contrast to our stereotypes, some of these metrics also involve other agents' opinions). For instance, two of the metrics proposed by~\cite{wu2010reputation}: the transaction price and the savviness of participants (measured as the focus of a participant on a particular category) are analogous to our stereotypes.


StereoTrust also has parallels to the ranking mechanisms used in web search engines. The usage of group information is analogous to using the content of the web pages to rank them. Transitive trust models resemble ``pure'' PageRank \cite{pagerank}, that uses only links between pages. Similarly to web search, where using both content and links together gives better results, we derive an enhanced method (d-StereoTrust), that uses both groups and (limited) trust transitivity.

\section{Model}
\label{preliminary}

We refer to a participant in the system as an agent. We denote by $\mathcal{A}$ the set of all agents in the system; and by $\mathcal{A}_x$ the set of agents known to agent $a_x$. An agent can provide services for other agents. A transaction in the system happens when an agent accepts another agent's service. To indicate the quality of a service, an agent can rank the transaction. For simplicity, we assume that the result is binary, i.e., successful or unsuccessful. $\Theta_{a_{x},a_{y}}$ denotes the set of transactions between service provision agent $a_{y}$ and service consumption agent $a_{x}$ and
$\theta_{a_x, a_y} = |\Theta_{a_{x},a_{y}}|$ denotes the number of such transactions.

We assume that each agent $x$ is characterized by a feature vector $f^x = [ f_1, f_2, \ldots, f_m]$. We denote by $F_i$ the set of values for each feature $f_i$,  $f_i \in F_i$. Features may be qualitative (gender, country of residence), or quantitative (salary, age, number of children). We assume that the domain of the feature vectors is the same for all agents. Transactions performed by agents are also characterized by similarly defined feature vectors.


\subsection{Group Definition}
\label{sec:group}
In StereoTrust, a group is a community of agents who show some common properties or behave similarly in certain aspects. Because of the common properties shared by all the members of a group, the group can act as a collective entity to represent its member agents (to a certain extent). For instance, people may
consider a programmer working in a well-known software company as skilled, even if they do not personally know the person. People trust the company based on the quality of produced software; thus they also trust the programmers who create the software. On the other hand, a company employing skilled (i.e., trusted) programmers can release high-quality products, and thus gain high reputation. Such interplay between the group's and its members' reputation is the basis of our work. We derive the trust of an agent according to the trusts of its corresponding groups.

Groups are defined subjectively by the trustor $a_x$. A group $G_x^i$ is a set of agents. We denote by $\mathcal{G}_x = \{G_{x}^1, G_{x}^2, \ldots, G_{x}^n \}$ the set of all groups defined by $a_x$. Based on $a_x$'s
previous experience, stereotypes, and any other information, $a_x$ formulates grouping function $M_x(G_x^i, a): [\mathcal{G}_x, \mathcal{A}_x] \to [0,1]$, that, for each group $G_x^i$, map agent $a$ to the probability that $a$ is the member of this group. Thus, in the most general model, a group is a fuzzy set of agents.  If $M_x(G_x^i, a) = 1$, it is certain that $a$ is member of $G_x^i$ ($a \in G_x^i$); if $M_x(G_x^i, a) = 0$, it is certain that $a$ does not belong to $G_x^i$ ($a \notin G_x^i$). Such a group definition makes the agent grouping flexible and personalizable, in contrast to the rigid notion of group membership in REGRET~\cite{regret}.


\subsection{Feature Selection}
\label{sec:featureselection}
The premise of our work is that the trustworthiness of a group (i.e., stereotypes) reflects the trustworthiness of the group members. Members of a group should behave consistently---ideally, a group should be discriminating, i.e. contain only either honest (the ones act honestly in transactions), or dishonest agents. Thus, the key question is: what features of agents can be used to form such discriminating groups?

Among many possible approaches to find discriminating features (such as the Gini index \cite{gini}, or the chi-square test \cite{chaid}) we use the information gain~\cite{informationtheory}, which is measured by \emph{entropy} as the criterion to select the features. The resulting decision model is easy to interpret by the users of the system. However, other, more complex approaches to derive decision criteria can be also used, such as decision trees or learning discriminant analysis (used in our recent work~\cite{metatrust,metatrustjournal}).

We assume that the categorization is binary (i.e., an agent is honest or dishonest).
We denote the proportion of honest and dishonest agents by $p_{h}$ and $p_{d}$ respectively ($p_d = 1-p_h$). Then the entropy of the set of agents that trustor $a_{x}$ has interacted with is calculated as:
\begin{equation}
   \label{eq:entropy}
   Entropy(\mathcal{A}_{x}) = -p_{h}log_{2}(p_{h}) - p_{d}log_{2}(p_{d}) \ .
\end{equation}

Entropy is used to characterize the (im)purity of a collection of examples. From Eq. \ref{eq:entropy} we can see that the entropy will have the minimum value of 0 when all agents belong to one class (i.e., either all honest or all dishonest) and the maximum value of $log_{2}2 = 1$ when agents are evenly distributed across the two classes. Using entropy, we calculate information gain of every feature to determine the best ones.

For each feature $F_i$, we partition the set of agents $\mathcal{A}_x$ into subsets $\mathcal{A}_{x,f_i}$ based on the values of the feature the agents have ($\mathcal{A}_{x,f_i} = \{ x : f^x_i = f_i \} $). The subsets $\mathcal{A}_{x,f_i}$ are disjoint and they cover $\mathcal{A}_{x}$.
The information gain of feature $F_{i}$ is then calculated by:

\begin{equation}
    \label{eq:informationgain}
    IGain(\mathcal{A}_{x} , F_{i}) = Entropy(\mathcal{A}_{x}) - \sum_{f_i}\frac{|A_{x,f_i}|}{|\mathcal{A}_{x}|}Entropy(\mathcal{A}_{x,f_{i}}) \ .
\end{equation}

The information gain of a feature measures expected reduction in entropy by considering this feature. Clearly, the higher the information gain, the lower the corresponding entropy. We can then select features based on their information gains. Several methods can be used. For instance, we can first rank the features by information gain (descending order) and choose the first $K$ features, depending on the specific applications; or we can set a threshold $\delta$ and select the features whose information gain is higher than $\delta$. We demonstrate in evaluation (see Section \ref{sec:evaluation}) how such feature selection scheme is applied in Epinions dataset to improve trust prediction accuracy.

In some cases, a trustor may also want to develop new stereotypes that are associated with combined features. For instance, consider that the agents' profile has many fields, including country, gender, income, etc. A trustor already has two stereotypes, say, on people from a country $A$ (with the feature vector $[ A, \cdot, \ldots, \cdot]$, where $\cdot$ denotes any value of the corresponding feature) and on women ($[ \cdot, \textrm{female}, \cdot, \ldots,  \cdot]$). When the trustor wants to develop a new stereotype on women from country $A$, it needs to combine the feature vectors. A combined feature vector contains all the values of qualitative features from vectors to be combined. Thus, in our example, the combined vector is $[ A, \textrm{female}, \cdot, \ldots, \cdot ]$. Note that contradictory feature vectors cannot be combined, for instance $[ \cdot, \textrm{female} , \cdot, \ldots, \cdot ]$ and $[\cdot, \textrm{male}, \cdot, \ldots, \cdot ]$. The trustor can transform other, non-qualitative features using standard machine learning techniques, such as discriminant analysis.

\subsection{Trust Model}
\label{sec:trustfunction}
While our approach is generic, in order to instantiate it, we use a computational trust model based on the beta distribution. A computational trust model models the complex notion of trust by a variable (binary, discrete, continuous, etc.). We assume that trust indicates the probability that an agent will perform a particular, expected action during a transaction~\cite{canwetrusttrust}. Thus, the agent's trust rating is a real number from range $[0,1]$, where 0 indicates that the agent is absolutely untrustworthy and 1 indicates that the agent is absolutely trustworthy.

The beta distribution is commonly used to model uncertainty about probability $p$ of a random event, including agent's reputation~\cite{betaRep02,Buchegger04arobust}. We model a series of transactions between a pair of agents as observations of independent Bernoulli trials. In each trial, the success probability $p$ is modeled by the beta distribution with parameters $\alpha$ and $\beta$ (we start with $\alpha=\beta=1$, that translate into complete uncertainty about the distribution of the parameter, modeled by the uniform distribution: $Beta(1,1)=U(0,1)$). After observing $s$ successes in $n$ trials, the posterior density of $p$ is $Beta(\alpha+s,\beta+n-s)$ \cite{bartoszynski2008probability}. We choose beta distribution because the resulting decision is easy to be interpreted by the system users, and is an already popular choice for modeling and interpreting trust.

The trust function between \emph{entities} $E_t$ (an individual agent or a group) is defined based on the beta distribution:

\begin{definition}[Trust Function]
Entity $E_{1}$ evaluates entity $E_{2}$. From the viewpoint of $E_{1}$, $s = S_{E_1,E_2}$ and $u = U_{E_1,E_2}$ represent, respectively, the number of successful transactions and unsuccessful transactions between $E_{1}$ and $E_{2}$ ($S_{E_1,E_2} \geq 0$
and $U_{E_1,E_2} \geq 0$). Trust function $T_{E_1,E_2}(p|s, u)$ mapping trust rating $p$ ($0 \leq p \leq 1$) to its probability is defined by:
\end{definition}
\begin{equation}
    \label{trustfunction}
    \begin{aligned}
    T_{E_1,E_2}(p | s,u) = \frac{(s+u+1)!}{s! u!} p^s(1 - p)^u \ .
    \end{aligned}
\end{equation}
The expected value of the trust function is equal to:
\begin{equation}
    \label{ev}
    E_{E_1,E_2}(T_{E_1,E_2}(p|s, u)) = \frac{s+1}{s+u+2} \ .
\end{equation}

\section{Basic StereoTrust Model}
\label{basicmodel}
The basic StereoTrust trust model only uses the trustor's local stereotypes to derive another agent's trustworthiness and hence it works without the target agent's behavioral history that is typically required by conventional trust models. Consider a scenario where a service requestor (a trustor agent) $a_{x}$ encounters a potential service provider $a_{y}$ with whom $a_{x}$ had no prior experience. We assume that $a_{x}$ can obtain some features about $a_{y}$, such as $a_{y}$'s interests, location, age etc.
$a_x$ combines its previous experience with such information to form groups that help to derive $a_y$'s trustworthiness.

StereoTrust starts by forming stereotypes, based on the trustor's historic information. In the first step, appropriate features are selected and/or combined (Section~\ref{sec:featureselection}). Using the processed features, StereoTrust groups agents accordingly. Note that StereoTrust considers only groups for which the membership is certain from $a_x$'s perspective ($M_x(G_{x,y}^i,a_y) = 1$). We denote these groups by  $\mathcal{G}_{x,y} = \{G_{x,y}^1, G_{x,y}^2, \ldots\}$ such that $M_x(G_{x,y}^i,a_y) = 1$ (for the sake of simplifying the notation, we will use $G^i$ in place of $G_{x,y}^i$ when the context is clear). The trust between $a_x$ and each of these groups $G^i$ is derived based on past interactions with agents that belong to $G^i$ with certainty. Thus, from the set of all agents $a_x$ has previously interacted with ($\mathcal{A}_x = \{ a_{1}, a_{2}, \ldots\}$), $a_x$ extracts those that belong to $G^i$ (i.e., $G^i = \{ a : M_x(G^i, a)=1 \}$). Then, $a_x$ counts the total number of successful $S_{a_x, G^i}$ transactions with $G^i$ as a sum of the numbers of successful transactions with $G^i$'s members: $S_{a_x, G^i}= \sum_{a \in G^i} S_{a_x,a}$.
The total number of unsuccessful transactions $U_{a_x,G^i}$ is computed similarly. In this way, $a_x$ can comprehensively understand how trustworthy is this group by aggregating its members' trustworthiness. Finally, $a_x$ uses Eq.~(\ref{trustfunction}) to derive $G^i$ trust function.

To derive agent $a_y$'s trust value, $a_x$ combines its trust towards all the groups $\mathcal{G}_x$ in which $a_y$ is a member. By combining multiple group trusts, $a_x$ is able to derive its trust in $a_y$ from multiple aspects.

The trust is computed as a weighted sum of groups' trust with weights proportional to the fraction of transactions with a group. For group $G^i$, weight factor $W_{x,y}^i$ is calculated as a number $\theta^i_{x,y}$ of $a_{x}$'s transactions with $G_{x,y}^i$ members ($\theta^i_{x,y} = | \Theta_{a_x,a} |$ such that $a \in G_{x,y}^i$); divided by the total number of  $a_x$'s transactions with members of any $\mathcal{G}_{x,y}$ group. Obviously, the higher the number of transactions regarding one group, the more likely is $a_{x}$ to interact with agents of this group, so this group contributes more to represent $a_{y}$'s trust from viewpoint of $a_{x}$.

We define weight factor $W^{i}_{x,y}$ for $G_{i}$ as:
\begin{equation}
    \label{L1W}
    W^{i}_{x,y} = \frac{\theta^i_{x,y}}{\sum_j \theta^j_{x,y}}
\end{equation}

\begin{figure}[tb]
\centerline{\includegraphics[scale=0.4]{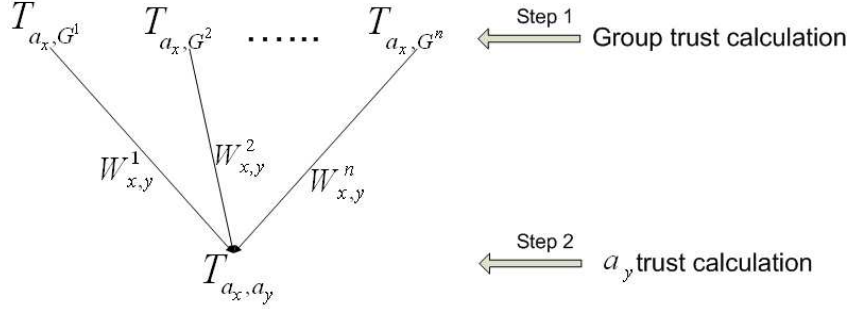}}
\caption{Process of trust calculation. Weighted sum of each group $G^{i}$'s
trust by assigning corresponding weight factor $W^{i}_{x,y}$} \label{fig:simplemodel}
\end{figure}

Using the estimated weights, we combine all group trusts to derive $a_{y}$'s trust. The process of trust calculation is illustrated in Figure \ref{fig:simplemodel}.

We propose two approaches to calculate and combine group trusts.
\subsubsection*{SOF Approach (Sum Of Functions)}
In this approach, we first calculate probability density of trust rating for each group using
trust function (Eq.~(\ref{trustfunction})) and then combine them to produce $a_{y}$'s probability density of
trust rating $TD_{a_x,a_y}(p)$ using Eq. (\ref{L1W}):
\begin{equation}
    \label{L1form1}
    TD_{a_x, a_y}(p) = \sum_{i} W^{i}_{x,y} \cdot T_{a_x,G^{i}}(p|S_{a_x, G^{i}},U_{a_x, G^i}) \textrm{,}
\end{equation}
where $S_{a_x, G^{i}}$ and $U_{a_x,G^{i}}$ are aggregated numbers of successful and unsuccessful transactions between $a_{x}$ and members of group $G^{i}$; $W^i_{x,y}$ is the weight (fraction of transactions with group $i$) and $T()$ is the resulting trust value.

\subsubsection*{SOP Approach (Sum Of Parameters)}
In this approach, we use only one trust function by setting the parameters, i.e., numbers of corresponding successful
and unsuccessful transactions.
\begin{equation}
    \label{L1form2}
    TD_{a_x,a_y}(p) = T_{a_x,a_y}(p|\sum_i W^{i}_{x,y} \cdot S_{a_x,G
^{i}}, \sum_i W^{i}_{x,y} \cdot U_{a_x, G^{i}}) \textrm{.}
\end{equation}

\section{Dichotomy Based Enhanced Model}
\label{sec:enhencedmodel}

\begin{figure}[tbp]
\vspace{-3mm} \centerline{\includegraphics[scale=0.3]{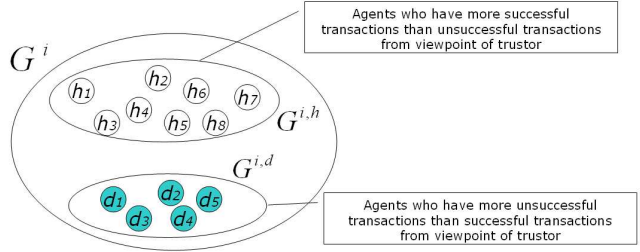}}
\vspace{-5mm} \caption{In d-StereoTrust, group $G^i$ is further divided into two subgroups containing exclusively ``honest'' and exclusively ``dishonest'' agents. When facing a stranger belonging to $G^i$, a trustor will estimate the similarity between the stranger and each subgroup.} \label{fig:IMgroup}
\end{figure}

\begin{figure}[tbp]
\vspace{-3mm} \centerline{\includegraphics[scale=0.475]{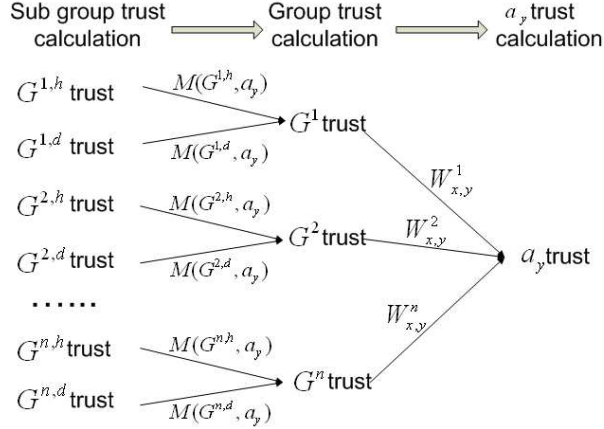}}
\vspace{-5mm} \caption{Process of trust calculation. Trusts of honest subgroup $G^{i,h}$ and dishonest sub
group $G^{i,d}$ of each group $G^{i}$ are firstly combined using closeness and then trusts
of all groups $G^{i}$ are combined using weight factor $W_{x,y}^i$ to derive target agent's trust.} \label{fig:improvedmodel}
\vspace{-1mm}
\end{figure}

StereoTrust model simply groups agents based on agents' and transactions' profiles. In some scenarios, it can be difficult for StereoTrust to accurately predict the performance of an agent who behaves differently from other members of its group(s). For instance, consider a case when trustor $a_x$ has interacted with mostly honest agents, while the target agent is malicious. StereoTrust will derive high trust for the malicious target agent.

To improve prediction accuracy, we propose dichotomy-based enhancement of StereoTrust (called d-StereoTrust). The key idea is to construct subgroups that divide agents on a finer level than the groups.
In d-StereoTrust, each top-level group $G^{i}$ is further divided into two subgroups, an \emph{honest} $G^{i,h}$ and a \emph{dishonest} $G^{i,d}$ subgroup (hence dichotomy-based). $a_x$ assigns an agent $a \in G^i$ to either subgroup by analyzing history of its transactions with $a$. The basic criterion we use is that if $a_x$ has more successful than unsuccessful transactions with $a$, $a$ is added to the honest subgroup $G^{i,h}$ (and, consequently, in the alternative case $a$ is added to $G^{i,d}$). Several alternative criteria are possible, for instance, the average rating of transactions with $a$. Note that, although not completely accurate, such a method indeed helps to separate honest agents from the dishonest ones more accurately than the basic StereoTrust.

After dividing a group $G^i$ into subgroups ($G^{i,h}$, $G^{i,d}$) and determining $a_x$'s trust towards the subgroups (computed as in the previous section), d-StereoTrust computes how similar is the target agent $a_y$ to the honest and the dishonest subgroup. If $a_y$ ``seems'' more honest, $a_x$'s trust towards aggregated $G^i$ should reflect more $a_x$'s trust towards the honest subgroup $G^{i,h}$; similarly, if $a_y$ ``seems'' more dishonest, the dishonest subgroup $G^{i,d}$ should have more impact on $a_x$'s aggregated trust towards $G^i$. This process is illustrated on Figure~\ref{fig:improvedmodel}.

The closeness, which can be measured by membership $M_x(G^{i,\cdot},a_y)$ of a target agent $a_y$ to subgroup $G^{i,\cdot}$ (where $\cdot$ represents $d$ or $h$) is based on other agents' opinions about $a_y$. Note that we cannot assign $a_y$ to a group (similarly to any other agent $a \in \mathcal{A}_x$), because the grouping described above is based on $a_x$ history with $a$, and, obviously, there are no previous transactions between $a_x$ and $a_y$. Thus, both $M_x(G^{i,h},a_y)$ and $M(G^{i,d},a_y)$ are fuzzy (in $[0,1]$).

An agent $a_x$ obtains opinions about a target agent $a_y$ by requesting a certain metric from other agents. For instance, $a_x$ can ask other agent $a_k$ about the percentage (denoted by $m_{k,y}$) of successful transactions it had with $a_y$. $a_x$ will seek opinions from honest agents from group $G^{i,h}$; and also from agents interested in $G^i$, but with no transactions with $a_x$ (based on their profile information, these agents could be classified as members of $G^i$, but they had no transactions with $a_x$). Obviously, the agents who have no transactions with $a_x$ may be dishonest thus may provide false reports.

Note that the amount of historic information needed from other agents in d-StereoTrust is a small subset of information required in models based on feedbacks or transitive trust. To collect feedbacks or to form transitive trust paths, Eigentrust-like algorithms must explore the whole network (that is take into account all the available historic transactions). In contrast, in d-StereoTrust, $a_x$ only asks the agents (from $a_x$'s perspective) from the honest sub-groups.

Based on all opinions $m_{k,y}$ received, $a_x$ computes an aggregated opinion $m_y$, which is used to measure the closeness of $a_y$ to subgroups as a simple average of $m_{k,y}$.

To characterize subgroups in a similar way, $a_x$ computes similar aggregation of its opinions towards subgroups $G^{i,h}$ and $G^{i,d}$. Aggregated opinion $m_{h}$ about subgroup $G^{i,h}$ is equal to the simple average of $m_{x,j}$ (percentage of successful transactions $a_x$ had with $a_j$), where $j$ is the index of agent $a_j \in G^{i,h}$. The aggregated opinion $m_{d}$ about the subgroup $G^{i,d}$ is derived in the same way.

Finally, the closeness between $a_y$ and each of the subgroups is computed as the fraction of the distance between $m_y$ from one side and $m_h$ and $m_d$ from the other:
\begin{equation}
    \label{L2Wh}
    M_x(G^{i,h},a_y) = \frac{1/(|m_y - m_h |)}{1/(|m_y - m_h |) + 1/(|m_y - m_d |)}
\end{equation}
\vspace{-2.5mm}
\begin{equation}
    \label{L2Wd}
    M_x(G^{i,d},a_y) = \frac{1/(|m_y - m_d |)}{1/(|m_y - m_h |) + 1/(|m_y - m_d |)}
\end{equation}

This procedure has a straightforward interpretation. If other agents have similar opinions about the target agent $a_y$ as $a_x$ has about the dishonest subgroup, then the target agent is most likely dishonest, so the dishonest subgroup trust should more influence $a_y$'s trust in the context of group $G^i$. Similarly, if other agents have experienced similar performance with $a_y$ as $a_x$ have with the honest group, then $a_y$ is most likely honest.

Note that we do not use the opinions provided by other entities to directly calculate $a_{y}$'s trust. Instead, we use them as metrics to measure the closeness between $a_{y}$ and the subgroups. In other words, we do not ask other agents ``is $a_y$ honest?'', but rather we ask about quantitative measures of experience they had with $a_y$ (for instance --- the percentage of successful transactions). This allows us, firstly, to be more objective; and, secondly, to easily extend d-StereoTrust to use multiple metrics taking into account different measures of satisfaction from a transaction (like the quality of service, the delivery time, etc.).


Also note that when other agents' opinions are not available, the d-StereoTrust model degrades to the StereoTrust model.

After calculating closeness, we combine groups' trusts to derive $a_{y}$'s trust. Similarly to the original StereoTrust, there are two approaches to combine various trust sources.

\subsubsection*{SOF Approach (Sum Of Functions)}
Using Eq.~(\ref{trustfunction})(\ref{L1W})(\ref{L2Wh}) and (\ref{L2Wd}) we have probability density
of target agent ($a_{y}$)'s trust rating $TD_{a_x,a_{y}}(x)$:
\vspace{-2mm}
\begin{equation}
    \label{L2form1}
    \begin{aligned}
    TD_{a_{x},a_{y}}(p)  = \sum_{i} \bigg( W_{x,y}^{i} & \Big(M_x(G^{i,h},a_y) \cdot T_{a_x,G^{i,h}}(p|S_{a_x,G^{i,h}},
        U_{a_x,G^{i,h}}) \\ & + M_x(G^{i,d},a_y) \cdot T_{a_x,G^{i,d}}(p|S_{a_x,G^{i,d}},U_{a_x,G^{i,d}}) \Big) \bigg) \textrm{,}
    \end{aligned}
\end{equation}
Where $S_{a_x,G^{i,h}}$/$S_{a_x,G^{i,d}}$ and $U_{a_x,G_{i,h}}$/$U_{a_x,G^{i,d}}$ are aggregated numbers of
successful and unsuccessful transactions of
each member of $G^{i}$'s subgroup $G^{i,h}$/$G^{i,d}$ from viewpoint of agent $a_{x}$; $W^i_{x,y}$ is the weight (fraction of transactions with group $i$); $M$ is the closeness of the agent to the subgroup; and $T()$ is the resulting trust value.

\subsubsection*{SOP Approach (Sum Of Parameters)}
Using Eq.~(\ref{trustfunction})(\ref{L1W})(\ref{L2Wh}) and (\ref{L2Wd}) we have probability density
of agent $a_{y}$'s trust rating $TD_{a_x,a_{y}}(x)$:
\begin{equation}
    \label{L2form2}
    \begin{aligned}
    TD_{a_x,a_{y}}(p) = T_{a_x,a_y}(p|&\sum_{i}^{}W_{x,y}^{i} \cdot (M_x(G^{i,h},a_y) \cdot S_{a_x,G^{i,h}} +
     M_x(G^{i,d},a_y) \cdot S_{a_{x},G^{i,d}}) \\ & ,\sum_{i}^{}W_{x,y}^{i} \cdot (M_x(G^{i,h},a_y) \cdot U_{a_x,G^{i,h}} + M_x(G^{i,d},a_y) \cdot U_{a_x,G^{i,d}}) \textrm{.}
    \end{aligned}
\end{equation}

\section{Trustworthy sharing of local knowledge}
\label{sec:share}

StereoTrust assumes that the trustor has sufficient local knowledge using which appropriate stereotypes can be efficiently formed. However, in some cases, agents' local knowledge may be very limited (e.g., agents who have recently joined the system or who do not interact with others frequently). To adapt StereoTrust to such situations, we propose to let agents share their stereotypes. Inexperienced agents can use such shared stereotypes to complement their limited local knowledge. The agents are, moreover, pragmatic about the external (shared by someone else) stereotypes: the influence of an external stereotype is weighted by its accuracy observed in previous interactions.

We first investigate what stereotypes are accurate enough to be shared in Section~\ref{sec:accuratestereotypes}. Then we present how the stereotypes sharing overlay network (SSON) is constructed in Section \ref{sec:sson}. The issues of external stereotypes combination, and SSON update are discussed in Section \ref{sec:combiningstereotype} and \ref{sec:SSONupdate} respectively. Note that SSON provides sophisticated sharing mechanisms (by maintaining a dedicated overlay network) for exchanging, combining, and updating stereotypes, and is well evaluated in contrast to heuristics that need further studies under realistic settings~\cite{bootstrapStereotype}.

\subsection{Accurate stereotypes}
\label{sec:accuratestereotypes}
In StereoTrust, a trustor's stereotype on one group is formed by aggregating
individuals' trusts (i.e., based on past experience) in this group.
This raises an interesting question: \emph{what
stereotypes are accurate enough to be shared with other agents?}
For instance, an agent itself may have few interactions with agents from a specific group, and hence its stereotypes about that group may not be comprehensive to make good estimate about the group members' behavior. Thus, even if the agent has no malicious intent, if it shares with others a stereotype which is in fact wrong or inconclusive, that will be detrimental to the system.

To address this issue, we need to determine which local stereotypes are accurate by investigating the minimum number of
transactions $M_{min}$ needed by a trustor to be confident (with some pre-determined confidence level, e.g., $0.95$) about the collective
behavior of the group. We treat a group of agents as an entity,
so we model the numbers of successful and unsuccessful transactions between the trustor and the group as the aggregated numbers of successful and unsuccessful transactions between trustor and members of that group. Using statistical methods, such as the Chernoff bound \cite{Mui02acomputational}, the agent can then derive  $M_{min}$.
Consequently, if a trustor has at least $M_{min}$ transactions
with members of one group, it is confident about its current
stereotype towards that group. Such a stereotype can be thus shared with other agents.

The above mechanism deals with mitigating the effect of inadvertently sharing misleading information by well behaved agents. Agents deliberately sharing misleading information, or agents with very different perspectives may be dealt with using conventional mechanisms, e.g., using a blacklist or a whitelist; next, we describe a strategy based on a whitelist.

\subsection{Stereotypes sharing overlay network (SSON)}
\label{sec:sson}

\begin{figure}[tbp]
\centerline{\includegraphics[scale=0.5]{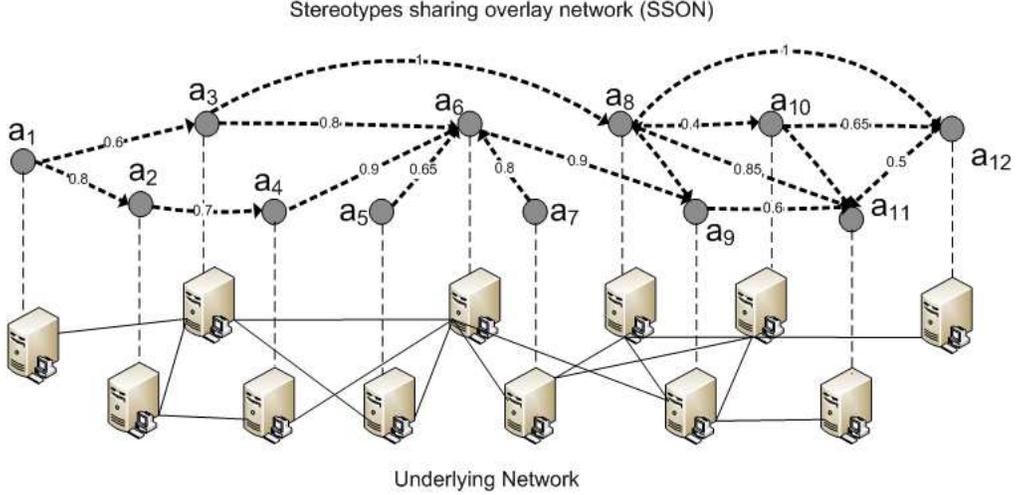}}
\caption{A stereotypes sharing overlay network.} \label{fig:overlay}
\end{figure}

To evaluate the trustworthiness of a target agent $a_{y}$, an inexperienced trustor $a_{x}$ who has few or no local knowledge needs to request other agents' stereotypes. $a_{x}$ only requests the agents who are trustworthy in terms of providing correct stereotypes. That is, each agent $a_{i}$ in the system maintains a \emph{Trusted Stereotype Provider} list $PROVIDER(a_{i})$ which stores the agents that this agent trusts in terms of providing correct local knowledge.
The stereotype sharing overlay network (SSON) is constructed by connecting agents and their trusted stereotype providers. SSON is a virtual network on top of current network infrastructure (e.g., a P2P network). Fig. \ref{fig:overlay} depicts a SSON, which is represented by a directed graph. The graph nodes correspond to the agents. The directed edges represent the trust relationship in terms of providing correct stereotypes. For instance, agent $a_{1}$ trusts agent $a_{2}$ and the corresponding edge (the trust relationship) is labeled with a trust score (i.e., 0.8 in this case). The trust score is determined and updated by the stereotype requester (i.e., $a_{x}$) after it evaluates correctness/usefulness of the information provided by the provider. Note that since an agent may have very subjective perspective, even if the provider provides accurate stereotype, it may not be correct/useful to $a_{x}$. In this case, the provider will be issued a low score although it is honest. In our work, the trust score falls into the range of $[0,1]$, where 1 represents a completely trustworthy provider and 0 represents a completely untrustworthy provider.

Initially, \emph{Trusted Stereotype Provider} list $PROVIDER(a_{x})$ is filled with $a_{x}$'s ``familiar'' agents (e.g., friends or colleagues in the real world, etc.). When no ``familiar'' agents exist in the system, $a_{x}$ chooses stereotype providers randomly. After each request, the trustor updates the trust score of the corresponding stereotype provider according to the accuracy of the reported stereotypes; the accuracy is assessed from the trustor's perspective (see Sec. \ref{sec:SSONupdate}). To collect correct stereotypes, $a_{x}$ request the top-K stereotype providers with highest trust scores in the \emph{Trusted Stereotype Provider} list, thus limiting communication overhead.

Notice that SSON is constructed to promote correct stereotypes sharing to help inexperienced agents estimate trustworthiness of an unknown service provider, but it is not used to discover trustworthy
transaction partner because agents who provide high quality service may not necessarily report correct information about other agents (and vice versa) \cite{Xiong04peertrust,Swamynathan05decouplingservice}.
In the scenario that the agents who provide correct stereotypes also act honestly in a transaction, SSON can be used to help promote successful transactions (i.e., select reliable agents who have high trust scores as the service providers). However, the goal of this work is to design mechanisms to estimate trustworthiness of an unknown service provider (i.e., no historical information is available), so discussion on relying SSON to derive agent's trust like traditional trust mechanisms (e.g., feedback aggregation \cite{betaRep02, p2prep, travos}, web of trust \cite{distributedtrust97, transitive03}, etc.) is out of the scope of this paper.

\subsubsection{Deriving trust by combining external stereotypes}
\label{sec:combiningstereotype}
After collecting other agents' stereotypes, we discuss how trustor $a_x$ combines these external stereotypes to derive the trust of the unknown target agent $a_y$.

We adopt a simple weight based strategy to combine the stereotypes in order to compute the final trust score for $a_y$. The weight of each stereotype depends on the trust (in terms of providing correct stereotype) of the corresponding stereotype provider.
We denote the trust scores of the stereotype providers by $T = \{t_{1}, t_{2}, t_{3},... \}$. The weight of stereotype provided by agent $a_{i}$ is calculated as $W_{i} = \frac{t_{i}}{\sum_{j}t_{j}}$.

Weighted stereotypes can be combined by the same methods as discussed in Section~\ref{basicmodel}: the SOF or the SOP methods.


\subsubsection{Updating the stereotype providers' trustworthiness}
\label{sec:SSONupdate}
The quality of stereotypes provided by different agents may vary. Some agents may maliciously provide fake information; others may have different perspective on the quality of transactions. An agent using external stereotypes must discover such behavior and update trust scores of the stereotype providers such that inaccurate stereotypes have less impact on the final decision.

The problem of deriving trust in a stereotype provider is analogous to the general problem of computational trust. $a_x$, after observing the outcome of its transaction with $a_y$, updates stereotype providers' trust. This time, however, we define a \emph{recommendation transaction} between a trustor $a_x$ and a stereotype provider $a_i$. The recommendation transaction is successful if the observed outcome of the original transaction (between a trustor $a_x$ and agent $a_y$) is the same as the outcome predicted by $a_i$'s stereotype.

The trust $t_i$ in stereotype provider $a_{i}$ can be then derived based on the number of successful $s_i$ and unsuccessful $u_i$ \emph{recommendation} transactions. $t_i$ is thus updated after each transaction.

Any computational trust model can be used to compute $t_i$ from $s_i$ and $u_i$. In order to instantiate the model, similarly to the basic StereoTrust model, we use the beta distribution (Section~\ref{sec:trustfunction}). $t_i$ is the expected value of the distribution, computed as $t_i = \frac{s_i+1}{s_i + u_i + 2}$.



\section{Discussion}
\label{sec:discussion}
So far, we have presented the basic StereoTrust model, its dichotomy based enhancement, as well as a overlay network supporting efficient stereotype information sharing.
The rationale for the StereoTrust approach is to determine an alternative and complementary mechanism to compute trust even in the absence of (global) information that is likely to be unavailable under some circumstances, and instead using some other class of information (stereotypes) which can be established by local interactions.
Since StereoTrust utilizes information theory/machine learning to conduct feature selection and feature combination to form stereotypes, it is worth mentioning that the trustor periodically re-trains the model to more accurately derive trustworthiness of the target agent. Several methods may be applied to update the model, for instance, the trustor may refine stereotypes after every $\tau$ new transactions; or the trustor only updates the model when the latest trust prediction is unsuccessful\footnote{Successful trust prediction indicates that current model is accurate enough so no update is needed.}. We will introduce and evaluate different update strategies in the experimental evaluation section (Section \ref{sec:evaluation}).

Similarly to other models like EigenTrust~\cite{eigen03}, Transitive Trust~\cite{transitive03}, BLADE~\cite{blade}, REGRET~\cite{regret} and Travos~\cite{travos}, StereoTrust is explicitly not designed to cope with agent's dynamic behavior. However, from the perspective of behavior science~\cite{MillerDriveReduction10}, as well as being supported by recent works that use contextual information to predict user behavior in various information systems~\cite{contextpredictcustomer,contextblogger,contextsensor2003}, we believe that an agent's behavior change in the transactions is correlated with and can be inferred (to certain extent) by the associated contextual information (e.g., by considering the dynamic trust \cite{trustpredictionijcai2011,hmmtrustaaai2012}). For instance, in an online auction site like eBay, a seller may vary his behavior consciously or unwittingly in selling different items (e.g., he may be careful when selling expensive goods, but imprudent with cheap ones).
By selecting appropriate features from the contextual information (e.g., the item's price), StereoTrust is able to construct stereotypes that partially model the target agents' implicit dynamic behaviors. Specifically, when an agent changes its behavior, some of the associated features may also vary accordingly. Such dynamism is observed by the trustor who will then adjust its local stereotypes to adapt to the target agent's dynamic behavior (we will show performance of such a learning based adaptive update strategy in the next section). Since this work focuses on ``cold-start'' problem of trust assessment, and handing dynamism is a problem in its own right, we leave
as a future work a more detailed discussion on addressing behavior changes.

\section{Evaluation}
\label{sec:evaluation}

In this section, we conduct experiments to evaluate the performance of proposed StereoTrust models.
We first discuss methodology in \ref{sec:method}. \ref{sec:realdata} presents the results on the Epinions.com dataset; \ref{sec:syndata} presents the results on the synthetic dataset.

\subsection{Methodology}
\label{sec:method}
When comparing Stereotrust with other algorithms and approaches, we consider two factors: the \emph{accuracy of prediction} that compares
the result of the algorithm with some ground truth; and the \emph{coverage}, the fraction
of the population that can be evaluated by the trustor, given trustor's limited knowledge.

We compare StereoTrust with the following algorithms.

\begin{description}
\item[\emph{Feedback Aggregation}]
    In this model \cite{betaRep02, Xiong04peertrust}, if the trustor does not know the target agent, it asks other agents and aggregate feedbacks to derive target agent's trust. Note that as the trustor may not have experience with the asked agent, it cannot identify the dishonest reporters, thus it may face false feedbacks.
\item[\emph{EigenTrust}]\cite{eigen03}
    This model uses transitivity of trust and aggregates trust from peers by having them perform a distributed calculation to derive eigenvector of the trust matrix. The trustor first queries its trusted agents (``friends'') about target agent's trustworthiness. Each opinion of a friend is weighted with the friend's global reputation. To get a broad view of the target agent's performance, the trustor will continue asking its friends' friends, and so on, until the difference of the two trust values derived in the two subsequent iterations is smaller than a predefined threshold. Pre-trusted agents (with high global reputation) are used in this model.
\item[\emph{Transitive Trust (Web of Trust)}]
    This model \cite{transitive03} is based on transitive trust chains. If the trustor doesn't know the target agent, it asks its trusted neighbors; the query is propagated until the target agent is eventually reached. The queries form a trust graph; two versions of the graph are commonly considered:
    \begin{description}
        \item[\emph{Shortest Path}]
        The agent chooses the shortest path (in terms of number of hops) and ignores the trustworthiness
        of agents along the path. If multiple shortest trust paths exist, the trustor will
        choose the most reliable one (the agents along the path are the most reliable).
        \item[\emph{Most Reliable Path}]
        The agent chooses the most reliable neighbor who has the highest
        trust rating to request for target agent's trust. If this neighbor does not know
        target agent, it continues requesting its own most reliable neighbor. To avoid long paths, the number of hops is limited to 6 in our experiment.
    \end{description}
\item[\emph{BLADE}]\cite{blade}.
  This approach models the target agent's properties (i.e., certain aspects of trustworthiness) and the feedback about the target agent collected from other agents as random variables. By establishing a correlation between the feedback and the target agent's properties using Bayesian learning approach (i.e., a conditional probability $Pr(R|F)$ where $R$ denotes the feedback and $F$ denotes a property), the trustor is able to infer the feedback provider's subjective feedback function to derive certain property of the unknown agent (e.g., does a seller ship correct goods on time in an online auction?). This avoids explicitly filtering out unreliable feedback. In other words, the trustor can safely use feedback from both honest and dishonest providers as long as they act consistently.
\item[\emph{Group Feedback Aggregation}]
    d-StereoTrust uses opinions reported by the agents who are the members of honest subgroups as the metrics to measure closeness between the target agent and the subgroups. We compare the accuracy of trust value derived using other agents' opinions (called group feedback aggregation) with that derived using d-StereoTrust to validate whether such third party information is used by d-StereoTrust judiciously. Note that different from the pure \emph{feedback aggregation} described above, \emph{group} feedback aggregation only uses the feedbacks provided by the agents from the honest subgroups.
\item[\emph{Dichotomy-only}]
  d-StereoTrust divides each group into an honest and a dishonest subgroups. To evaluate the impact of the initial grouping in d-StereoTrust, we compare d-StereoTrust with a similar, dichotomy-based algorithm, but without the higher-level grouping (thus---without the stereotypes). In dichotomy-only, agent $a_{x}$ classifies all the agents it has previously interacted with into just two groups: honest and dishonest (``honest agents'' having more successful transactions with $a_{x}$). Similarly to d-StereoTrust, to evaluate an agent $a_{y}$, $a_x$ queries the agents $a_i$ from the honest subgroup about their trust to $a_{y}$; using these feedbacks, $a_x$ calculates the distance between $a_{y}$ and the two groups.
\end{description}

To summarize, the feedback aggregation, the Eigentrust (and its variants) and the transitive trust model (actually the basis of Eigentrust) are currently the mainstream trust mechanisms.
BLADE, similarly to our approach, uses agent's features to determine trust. Group feedback aggregation and dichotomy-only are by-products of our approach; their results will
quantify the impact of each element of (d-)StereoTrust.


To estimate the accuracy of each algorithm, we compare the value of trust computed by the algorithm for a pair of agents with the ground truth. Then, we aggregate these differences over different pairs using the mean absolute error (MAE).

Besides prediction accuracy, we also measure the coverage of each algorithm; the coverage is defined as the percentage of agents in the system that can be evaluated by a trustor.

We present the accuracy of the algorithms in two formats. First, to measure overall performance of an algorithm, we show the MAE aggregated over the whole population of agents (e.g., Table~\ref{tab:epinionsMAE}). Second, to see how the algorithm performs in function of agent's true trustworthiness (the ground truth), we construct figures presenting the derived trust for a subset of agents (e.g., Figure~\ref{fig:epinionsm}).
Of course, as in the real system, the trust algorithms do not have access to the ground truth.
To avoid cluttering, we randomly choose 50 target agents. The y-axis represents the trust rating of the agents;  The x-axis represents the index of the evaluated agent. For clarity, agents are ordered by decreasing true trustworthiness.

Ideally, the ground truth of an agent represents the agent's objective trustworthiness. However, as we are not able to measure it directly, we have to estimate it using the available data. Along with the description of each dataset, we discuss how to derive the ground.

\subsection{Epinions Dataset}
\label{sec:realdata}
Epinions.com is a web site where users can write reviews about the products and services, such as books, cars, music, etc (later on we use a generic term ``product''). A review should give the reader detailed information about a specific product. Other users can rate the quality of the review by specifying whether it was \emph{Off Topic}, \emph{Not Helpful}, \emph{Somewhat Helpful}, \emph{Helpful}, \emph{Very Helpful} or \emph{Most Helpful}. For each review, Epinions.com shows an average score assigned by users.

Epinions.com structures products in tree-like categories. Each category (e.g., books) can include more specific categories (e.g., adventure, non-fiction, etc.).  The deeper the level, the more specific the category to which the product belongs.

The complete Epinions dataset we crawled contains 5,215 users, 224,500 reviews and  5,049,988 ratings of these reviews. For our experiments, we selected 20 trustors and 150 target agents randomly (we repeated the experiments with different agents and got similar results). On the result plots (e.g., Figure~\ref{fig:epinionsm}), error bars are added to show the 95\% confidence interval of predictions by each trustor for the same target agent

In Epinions community, users write or rate reviews of products they are interested in. This results in an intuitive grouping criterion: groups correspond to categories of products, and
an Epinions user belongs to a certain group if she wrote or rated a review of a product in the corresponding category.

\subsubsection{Modeling Epinions to the StereoTrust Model}

To map Epinions.com to StereoTrust model, we treat each user as an agent. Epinions.com categories provide a natural representation of the \emph{interested in} relation. A user is \emph{interested in} a (sub)category if she wrote or rated at least one review of a product under this category. Groups are formed according to agents' \emph{interested in} relations. Consequently, each Epinions.com category corresponds to a group of agents, each of whom is \emph{interested in} (wrote or rated a review for) this category. Note that if there exist multiple such categories (i.e., stereotypes), in order to improve trust prediction accuracy, we only select the first three ones that have the highest information gains (see Section \ref{sec:featureselection}).

A transaction between agents $a_x$ and $a_y$ occurs when $a_x$ rates a review written by $a_y$; the outcome of the transaction corresponds to the assigned rating. To map Epinions.com ratings to StereoTrust's binary transaction outcome, we assume that the transaction is successful only if the assigned score is \emph{Very Helpful} or \emph{Most Helpful}. We set the threshold so high to avoid extreme sparsity of unsuccessful transactions (over 91\% review ratings are \emph{Very Helpful} or \emph{Most Helpful}).

We compute the ground truth of an agent as the average rating of the reviews written by the agent. For instance,
if an agent wrote 3 reviews, the first review was ranked by two users as 0.75 and 1.0 respectively, while the second and the third received one ranking each (0.75 and 0.5), the ground truth for that user is  equal to $(0.75+1.0+0.75+0.5)/4$. Note that the ``ground truth'' computed with this simple method only approximates the real trustworthiness of an agent, as we do not adjust the scores to counteract, e.g., positive or negative biases of the scoring agents. In order to avoid biased ground truth caused by individual erroneous ratings, we removed the reviews with small amount of ratings (less than 50 ratings in our experiments).

\subsubsection{Results}

\begin{figure}[tbp]
  \begin{center}
    \centerline{\includegraphics[scale=0.32]{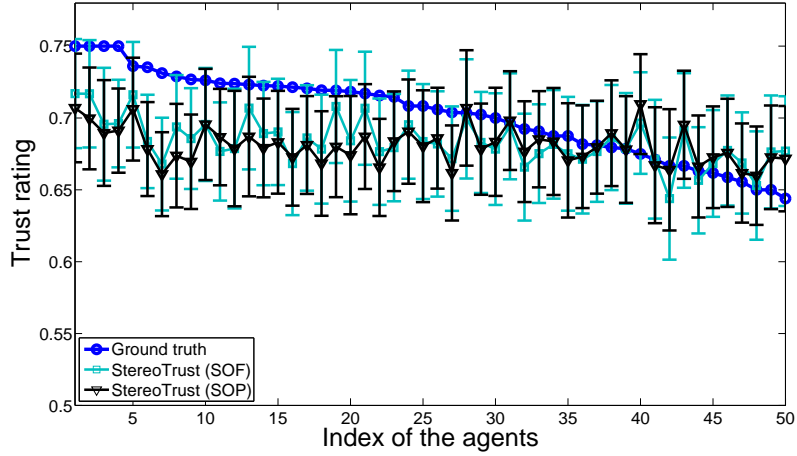}}
  \caption{Comparison of StereoTrust model and the ground
    truth on Epinions.com dataset.}
    \label{fig:epinionsm}
  \end{center}
\end{figure}

\begin{figure}[tbp]
  \begin{center}
    \centerline{\includegraphics[scale=0.32]{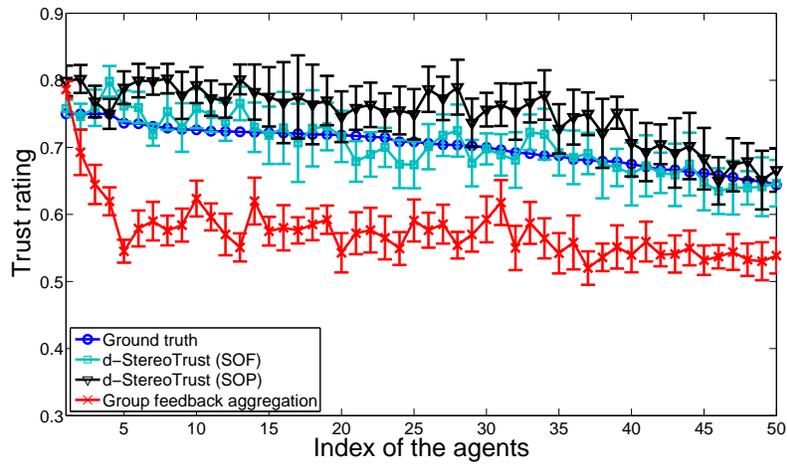}}
  \caption{Comparison of d-StereoTrust, group feedback aggregation and the ground
    truth on Epinions.com dataset.}
    \label{fig:epinionsim}
  \end{center}
\end{figure}

\begin{figure}[tbp]
  \begin{center}
    \centerline{\includegraphics[scale=0.32]{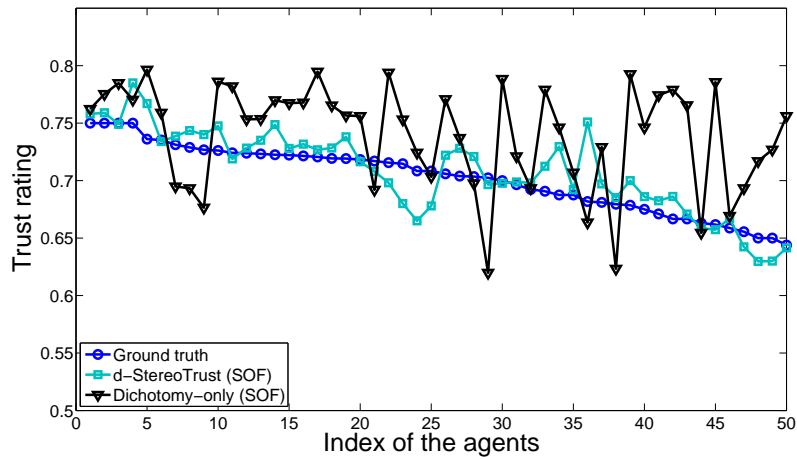}}
  \caption{Comparison of d-StereoTrust model
   and dichotomy-only using Epinions.com dataset.}
    \label{fig:compareallReal}
  \end{center}
\end{figure}

We select the top 3 features (i.e., categories\footnote{The real name of these categories can not be provided due to hashing (for privacy purpose).} in the experiments) based on information gain (see Section \ref{sec:featureselection}) for constructing stereotypes. Note that all figures in the evaluation section show results with feature selection.
Figure \ref{fig:epinionsm} shows the performance of StereoTrust model. SOF/SOP on the legend indicates that the stereotypes are aggregated using SOF/SOP approach respectively (Section~\ref{basicmodel}). The results show that  both SOF and SOP approaches fail to provide a good fit to the ground truth. This is because in the Epinions dataset, most ratings given by the agents are positive (\emph{Very Helpful} or \emph{Most Helpful}). As most of the trustors have limited direct experience with low quality reviews (hence -- dishonest agents), using only locally-derived stereotypes it is difficult to predict that an agent will write low quality reviews.

Figure \ref{fig:epinionsim} show the performance of d-StereoTrust model.
We can see that both SOF and SOP derived trust ratings are more accurate than feedbacks
derived trust rating (group feedback aggregation), which supports that our model outperforms the one which simply aggregates other
agents' feedbacks. SOF approach gives a better fit to
the ground truth than SOP approach. Comparing Figure \ref{fig:epinionsm} and \ref{fig:epinionsim},
we observe that d-StereoTrust results fit the ground truth better than the StereoTrust. Thus, as we expected, taking into account other agents' feedback improves prediction accuracy.

Figure \ref{fig:compareallReal} compares d-StereoTrust model with dichotomy-only (StereoTrust is omitted as it is worse than d-StereoTrust in terms of prediction accuracy). Error bars are removed for clarity and only SOF approach, which outperforms SOP approach is showed for each model. We see that d-StereoTrust model provides more accurate prediction than dichotomy-only does. This proves that by considering both stereotypes and global information we are able to predict target agent's behavior more accurately than using only the global information.

\begin{table}[tb]
\caption{Mean Absolute Error (with 95\% Confidence Interval). The values separated by `/' shows the results with (left) and without (right) feature selection respectively. Epinions.com dataset. SP -- shortest path; MRP -- most reliable path}\hspace{-1cm}
\label{tab:epinionsMAE}
\begin{tabular}{ l | c | c }
\hline
algorithm    &  MAE & 95\% C.I.\\
\hline
\hline
\bfseries d-StereoTrust (SOF) &  \bfseries 0.0512/\bfseries 0.0632 &\bfseries (0.0472,0.0561)/\bfseries (0.0586,0.0678) \\
d-StereoTrust (SOP) &  0.1105/0.1299 & (0.1056,0.1149)/(0.1245,0.1353)\\

StereoTrust (SOF) & 0.1006/0.1114  & (0.0945,0.1071)/(0.1067,0.1161)\\
StereoTrust (SOP) & 0.1012/0.1177 & (0.0968,0.1055)/(0.1136,0.1218)\\

Dichotomy-only (SOF) & 0.1245/0.1365 & (0.1198,0.1295)/(0.1307,0.1423)\\
Dichotomy-only (SOP) & 0.1526/0.1750 & (0.1475,0.1572)/(0.1690,0.1810)\\

Group feedback aggregation  &  0.1286/0.1452 & (0.1233,0.1341)/(0.1386,0.1518) \\
\hline
\end{tabular}
\end{table}

Table \ref{tab:epinionsMAE} lists the calculated the MAE along with 95\% confidence interval for all the trust mechanisms. In order to demonstrate performance improvement by the feature selection strategy, we show both results with feature selection (on the left of `/') and that without feature selection (on the right of `/').


\begin{figure}[tbh]
  \hspace{-5mm}
    \subfigure[\label{fig:update}Performance of different model update strategies.]
    {\includegraphics[height=5cm, width=7.5cm]{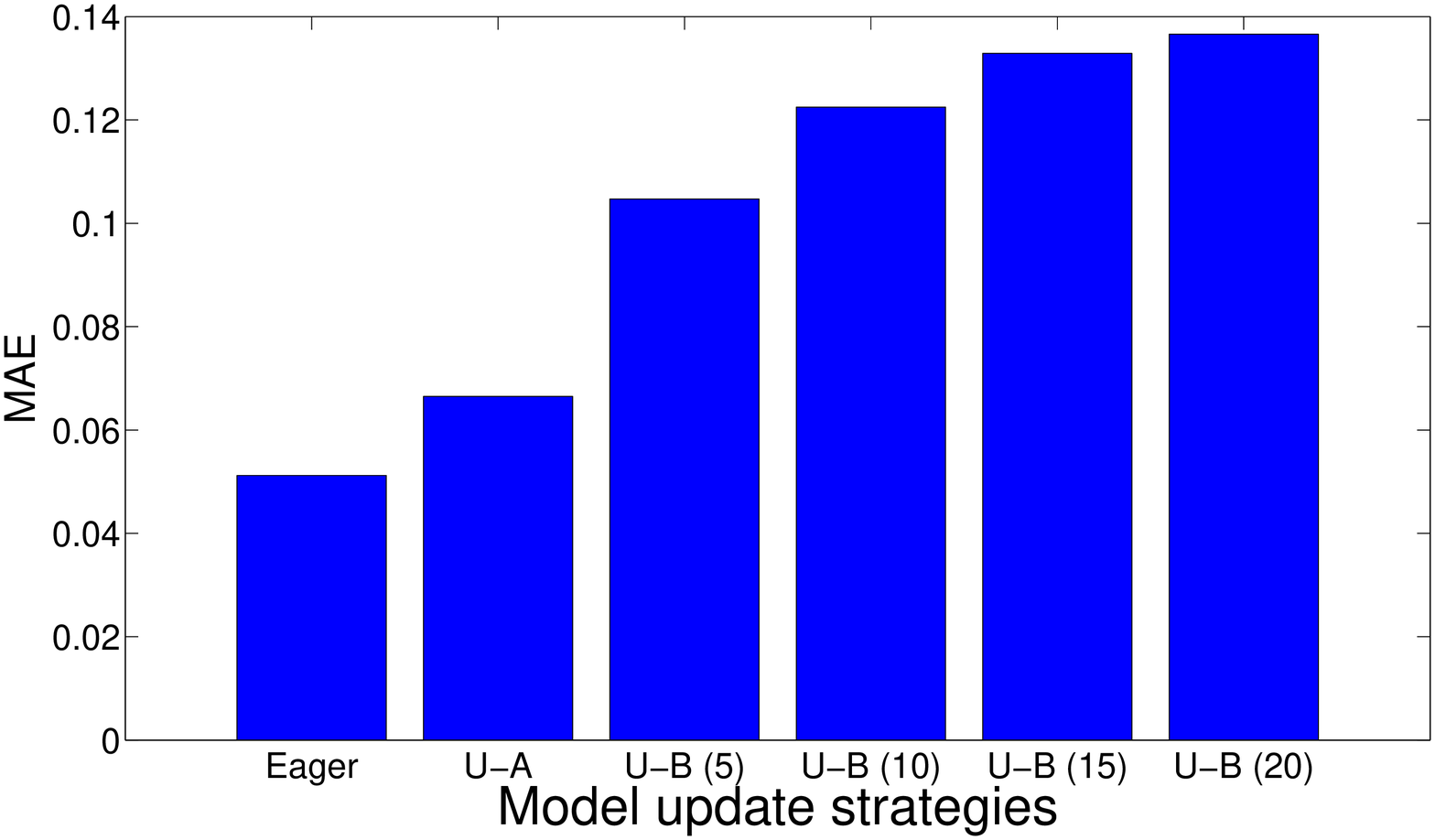}}\hspace{-0mm}
    \subfigure[\label{fig:updatecomplexity}Normalized cost of computation of different model update strategies.]{\includegraphics[height=5cm, width=7.5cm]{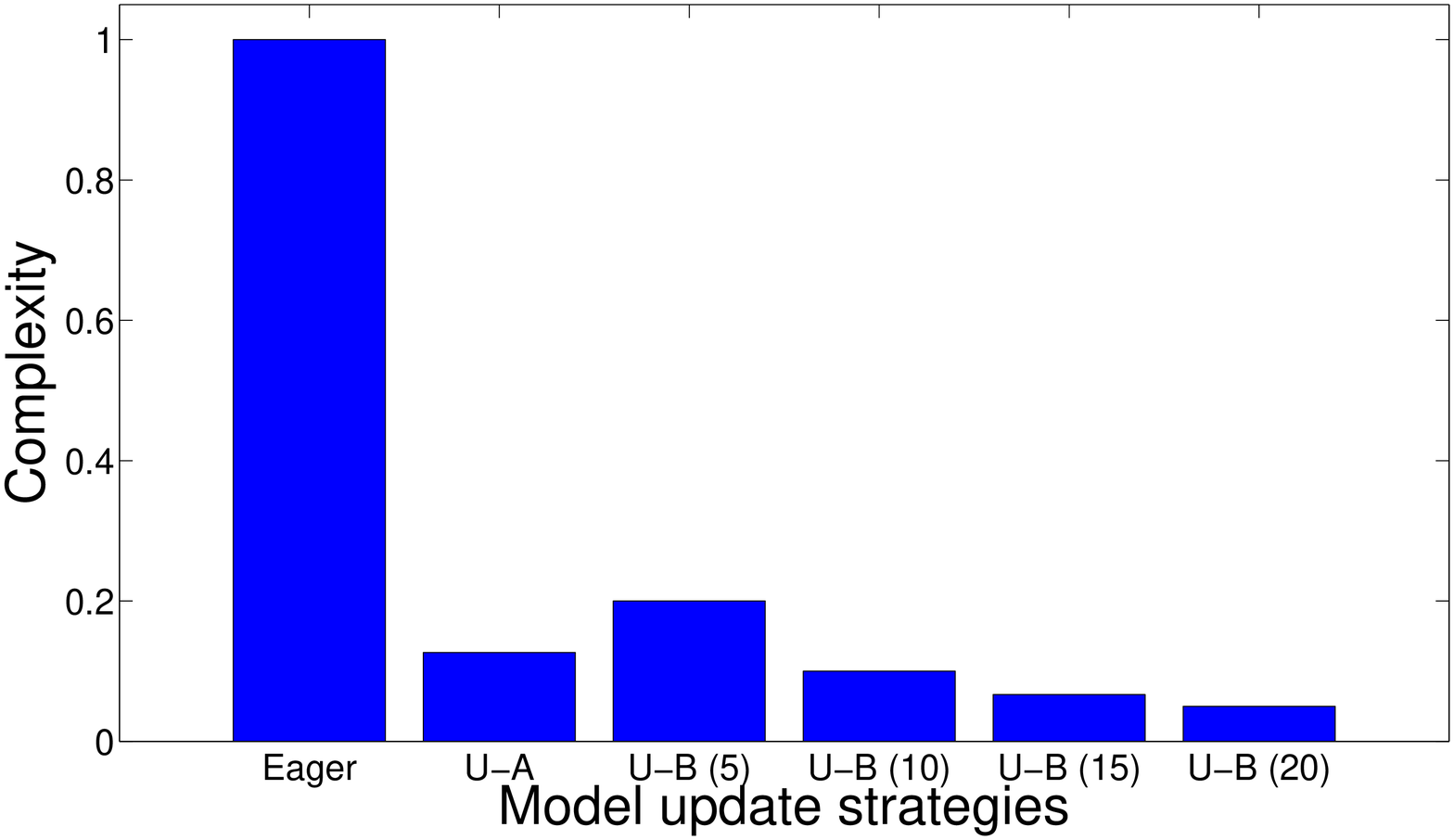}}
  \caption{Comparison of different model update strategies.}
\label{fig:updatestrategies}
\end{figure}

We also demonstrate how prediction accuracy of StereoTrust is influenced by different model update strategies (see Section \ref{sec:discussion}). Note that in the experiments so far, the trustor updates StereoTrust model after each new transaction (i.e., reconstructing the relevant stereotypes by considering outcome/features of the new transactions). Such frequent update ensures that StereoTrust model closely follows the agent's experience. However, frequent updates obviously incur computational overheads. We compare the \emph{eager} update strategy with two \emph{lazy} heuristics: (1) U-A, the trustor only updates the model when the latest trust prediction is unsuccessful; (2) U-B ($\tau$), the trustor updates the model after every $\tau$ transactions. Figure \ref{fig:update} shows the performance of StereoTrust (SOF) when different update strategies are applied. Note that on average, each target agent changes its behavior\footnote{Please differentiate this kind of dynamic behavior from another ``dishonest'' behavior, i.e., an agent writes high quality reviews in certain categories but low quality ones in other categories (please refer to the motivation example about Epinions.com in the Introduction).}. (i.e., writing low quality reviews in a category where it normally provides high quality reviews) with probability of around 10\%. Obviously, eager strategy, which takes into account the latest transaction information, outperforms other lazy strategies. U-A is worse than the eager strategy, but is better than U-B ($\tau$). This is because U-B ($\tau$) is ``agnostic'' to the dynamism of the agents' behavior (i.e., U-B ($\tau$) has to wait until $\tau$ new transactions happen), and hence has to spend very high (depending on the value of $\tau$) amount of update cost continuously, mostly without any utility. In contrast the adaptive strategy U-A risks having one erroneous decision but saves significantly on the efforts to keep the model reasonably accurate.
Figure \ref{fig:updatecomplexity} shows different model update strategies' computational complexities that are normalized to the range $[0,1]$. We conclude that a suitable update strategy can reduce computational overheads significantly while keeping the performance reasonably high.


\subsubsection{Discussion}
StereoTrust, using only stereotypes (category information)
to form groups, does not predict target agent's performance
accurately because Epinions.com is a friendly community. Users are likely to give high ratings to
reviews written by others; in consequence, successful transactions dominate. In such environment, StereoTrust does not have enough negative experience to form negative stereotypes, and thus overestimates trust. Due to the same reason, dichotomy-only does
not work well either. In contrast, d-StereoTrust improves the prediction
accuracy. This proves that both elements of our algorithms --- stereotypes (groups) and third-party information --- are crucial for an accurate prediction. In all cases, feature selection helps to improve the performance of various StereoTrust variants.

\subsection{Synthetic Dataset}
\label{sec:syndata}
Epinions.com has a friendly community with few dishonest agents. To test StereoTrust in a more hostile environment, we generated a synthetic dataset simulating a hostile version of Epinions.com-like community.

\subsubsection{Synthetic Dataset Generation}

The synthetic dataset contains 200 agents; 40\% of them are dishonest. A honest agent provides a high quality review (with true quality = 0.6 or 0.8 or 1.0) or a true feedback\footnote{A true feedback accurately reflects the provider's opinion that matches the quality of a review, and false one does not.} with a probability $P_m$. A dishonest agent provides a low quality review (with true quality = 0.0, 0.2, 0.4) or a false feedback with the same probability. Note that we tried different possibility values $P_m \in \{0.6, 0.7, 0.8, 0.9\}$ and observed the same qualitative trends. Consider the informativeness of the evaluation results, we just report one set of the results, i.e., $P_m = 0.9$. If a true feedback has a value of $\lambda$, the corresponding false feedback is a value of
$1 - \lambda$. Both the number of reviews written by an agent and the number of ratings of a review are generated by a normal distribution ($\mu=10$, $\sigma=4$). The agents who assign ratings to a review are selected randomly from a set of agents who are also interested in the review's category.

We simulate an environment with 12 categories (indexed $1,2,..., 12$) and 20 products in each category. Honest and dishonest agents are biased towards different categories.
A honest agent with probability $0.7$ writes a review for a product from categories $1,2,3,4$;  with probability $0.21$ for products from categories $9,10,11,12$; and with probability $0.03$ for products from categories $5,6,7,8$. A dishonest agent with probability $0.7$ writes a review for products from categories $5,6,7,8$; with probability $0.21$ for products from categories $9,10,11,12$; and with probability $0.03$ for products from categories $1,2,3,4$. Note that for a trustor, if there exist multiple such categories (i.e., stereotypes), in order to improve trust prediction accuracy, it only selects the first three ones that have the highest information gains (see Section \ref{sec:featureselection}).

We compute the ground truth of an agent as the average rating of the reviews written by this agent. Different from the ground truth in the Epinions dataset, in the synthetic dataset, rating of one review is determined by the design of the dataset, so this rating represents the real quality of the review, thus
the calculated ground truth better approximates the objective trustworthiness of an agent.

We choose one honest agent randomly as a trustor and we predict behavior of other agents in the system. Each experiment is repeated 10 times. Each run uses a different synthetic dataset. In the figures, error bars represent 95\% confidence interval.

\subsubsection{Results}

\begin{figure}[tbp]
  \begin{center}
    \centerline{\includegraphics[scale=0.32]{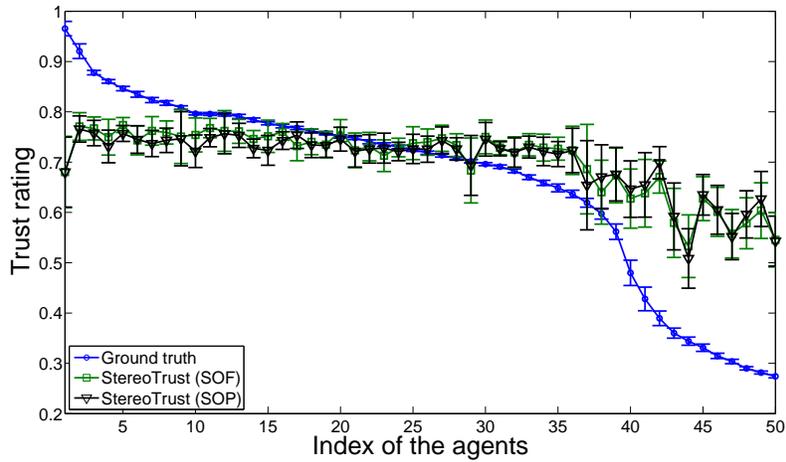}}
  \caption{Comparison of the basic StereoTrust and the ground
    truth on the synthetic dataset.}
    \label{fig:synsm}
  \end{center}
\end
{figure}

\begin{figure}[tbp]
  \begin{center}
    \centerline{\includegraphics[scale=0.32]{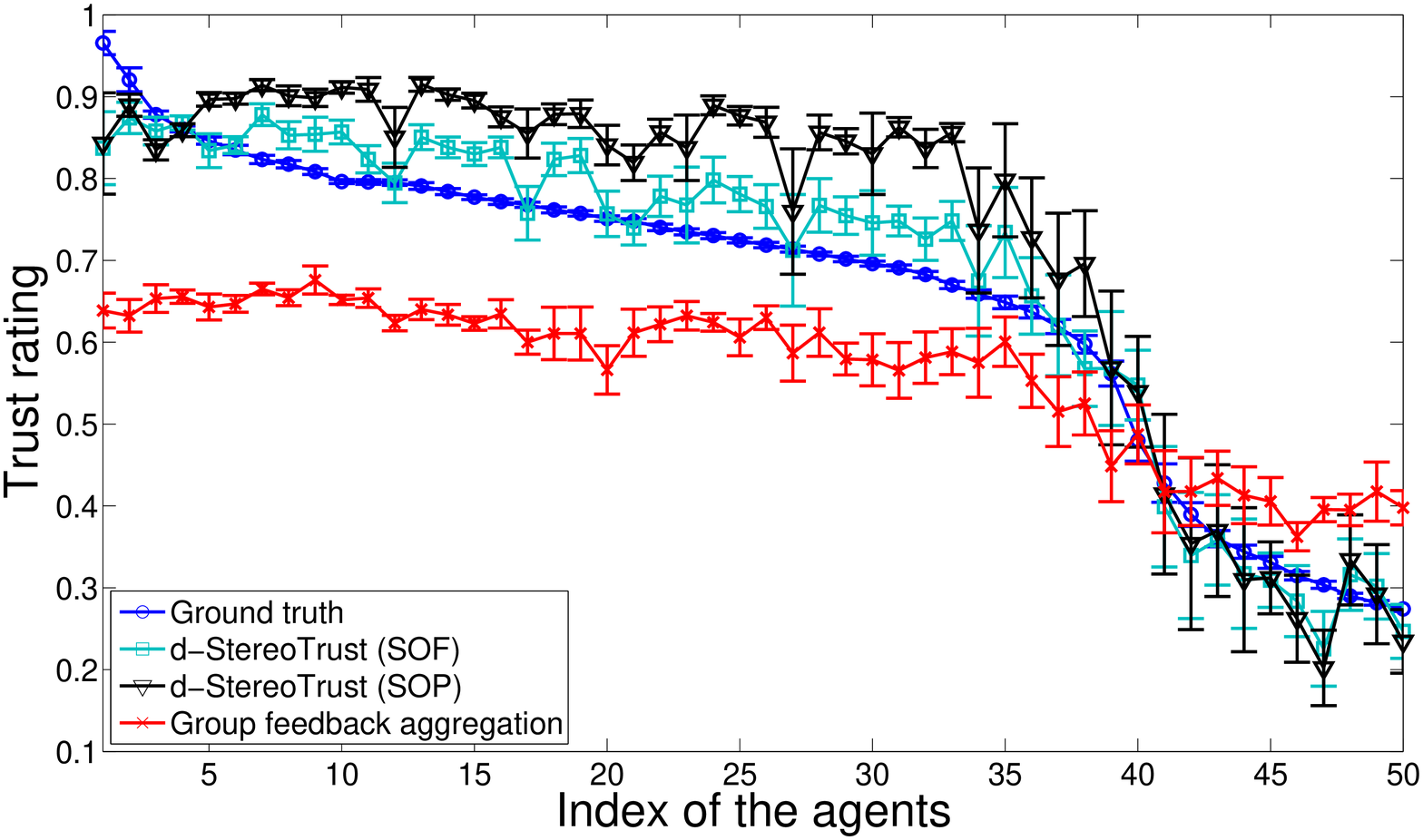}}
  \caption{Comparison of the d-StereoTrust and the ground
    truth on the synthetic dataset.}
    \label{fig:synim}
  \end{center}
\end{figure}

\begin{figure}[tb]
  \begin{center}
    \centerline{\includegraphics[scale=0.32]{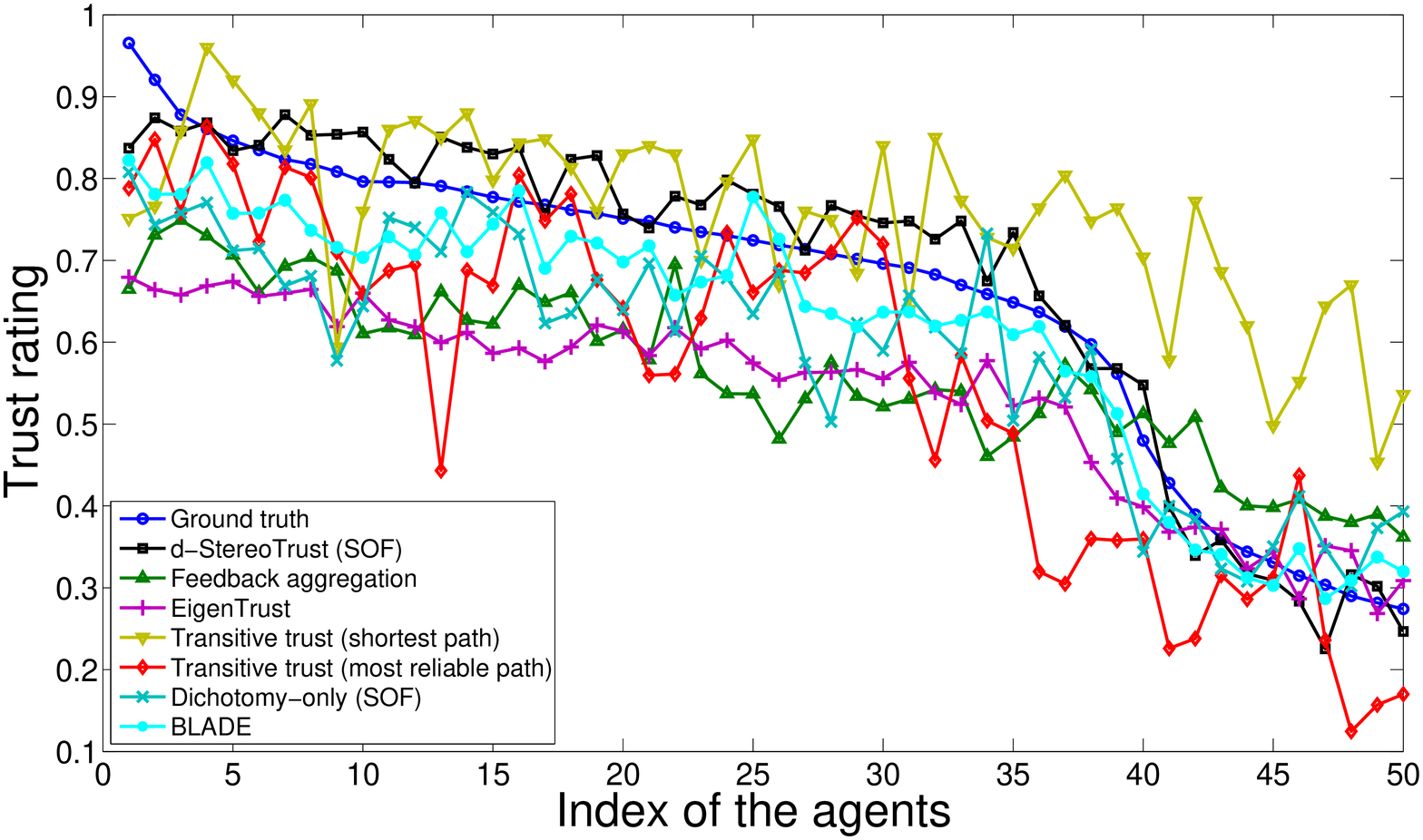}}
  \caption{Comparison of all the algorithms on the synthetic dataset.}
    \label{fig:synall}
  \end{center}
\end{figure}

Similarly to the Epinions dataset, we select the top-3 features (categories) to build stereotypes.
We first demonstrate evaluation results without using stereotype sharing overlay network (SSON), i.e., the trustor has sufficient local knowledge. Figure \ref{fig:synsm} shows the performance of the basic StereoTrust model. The trust derived by the basic StereoTrust fits the ground truth in general, but the match is not very close (see Table \ref{tab:syndataMAE} for numerical results). However, the trend looks better than that in the Epinions dataset.

Figure \ref{fig:synim} shows the performance of d-StereoTrust. Obviously, d-StereoTrust provides more accurate prediction than the StereoTrust. This is because d-StereoTrust forms groups having finer granularity; thus the local trust information and the third party information are properly used to represent target agent's trust. Similarly to the Epinions dataset, the group feedback aggregation is less accurate than SOF/SOP.

Figure \ref{fig:synall} compares d-StereoTrust model (using SOF) with dichotomy-only (using SOF) and other algorithms. Error bars are removed for clarity.  We observe that the existing algorithms predict target agent's trust less accurately
than d-StereoTrust does. These existing algorithms show obvious gaps between the ground truth and the derived rating either for the honest target agents (EigenTrust and the most reliable path transitive trust) or
for the dishonest target agents (shortest path transitive trust); or for all target agents (feedback aggregation).

\begin{table}[tb]
\caption{Predicting the true trust (the true review quality) in the synthetic dataset. Mean absolute error and coverage. Confidence intervals computed over 10 repetitions. SP -- shortest path; MRP -- most reliable path.}\hspace{-1.8cm}
\label{tab:syndataMAE}
\begin{tabular}{ l | c | c | c |c }
\hline
algorithm & honest agents & dishonest agents & all agents (95\% C.I.) & coverage \\
\hline
\bfseries d-StereoTrust (SOF) & \bfseries 0.1046 &  0.0968 & \bfseries 0.1006 (0.0968,0.1047) &  95.5\% \\
EigenTrust & 0.1487 & \bfseries 0.1002 & 0.1263 (0.0966,0.1510) & 96.4\% \\
BLADE & 0.1305 & 0.1109 & 0.1215 (0.1012,0.1395) & 98.5\% \\
Dichotomy-only (SOF) & 0.1306 & 0.1205 & 0.1248 (0.1202,0.1374) & 96.3\% \\
StereoTrust (SOF) & 0.1307 & 0.2588 & 0.1790 (0.1296,0.2348) & 96.9\% \\
Feedback aggregation & 0.1450 & 0.1642 & 0.1535 (0.1432,0.1678) & \bfseries 99.9\% \\
Transitive trust (SP) & 0.1547 & 0.3319 & 0.2304 (0.1424,0.3384) & 99.3\% \\
Transitive trust (MRP) & 0.1468 & 0.1678 & 0.1552 (0.1416,0.1688) & 82.1\% \\
\hline
\end{tabular}
\end{table}

Table \ref{tab:syndataMAE} summaries the MAE (with 95\% confidence interval for all the agents) and coverage of each model involved in comparison. For each model, we show the mean absolute error (MAE) for evaluation of the honest target agents, of the dishonest target agents and of all the target agents (we distinguish between these groups, as typically the accuracy on the larger group is better -- but it is the accuracy of detecting dishonest agents that usually should be optimized). Note that for StereoTrust, d-StereoTrust and dichotomy-only, we show SOF results only (as it constantly outperforms SOP). For d-StereoTrust, EigenTrust, Dichotomy-only and BLADE, although the confidence intervals overlap, the resulting differences are statistically significant (two-tailed, paired t-test, p-values $<0.001$). From this table, we observe that by reasonably combining the trustor's local stereotypes and small amount of third party knowledge, d-StereoTrust outperforms other trust models.

\begin{table}[tb]
\caption{Prediction of the true trust (the true review quality) in the synthetic dataset. The values separated by `/' shows the results with (left) and without (right) Stereotype-Sharing Overlay Network (SSON). Mean absolute error; confidence intervals computed over 10 repetitions}\hspace{-1.5cm}
\label{tab:syndataMAESSON}
\begin{tabular}{ l | c | c | c}
\hline
trustor & honest agents & dishonest agents & all agents (with 95\% C.I.)  \\
\hline
agent 1 &  0.1273/0.1485 & 0.1196/0.1368 & 0.1221 (0.1130,0.1314)/0.1428 (0.1345,0.1502)\\
agent 2 &  0.1261/0.1491 & 0.1212/0.1377 & 0.1232 (0.1142,0.1323)/0.1441 (0.1352,0.1510)\\
agent 3 &  0.1291/0.1488 & 0.1204/0.1362 & 0.1253 (0.1163,0.1330)/0.1434 (0.1341,0.1502)\\
agent 4 &  0.1292/0.1471 & 0.1202/0.1358 & 0.1254 (0.1162,0.1345)/0.1429 (0.1359,0.1497)\\
agent 5 &  0.1286/0.1475 & 0.1201/0.1367 & 0.1242 (0.1149,0.1332)/0.1428 (0.1366,0.1511)\\
agent 6 &  0.1268/0.1469 & 0.1215/0.1359 & 0.1239 (0.1145,0.1331)/0.1420 (0.1351,0.1493)\\
agent 7 &  0.1288/0.1492 & 0.1231/0.1389 & 0.1251 (0.1124,0.1338)/0.1442 (0.1365,0.1516)\\
agent 8 &  0.1275/0.1475 & 0.1223/0.1372 & 0.1248 (0.1158,0.1337)/0.1425 (0.1347,0.1504)\\
agent 9 &  0.1268/0.1481 & 0.1218/0.1376 & 0.1241 (0.1152,0.1326)/0.1431 (0.1353,0.1511)\\
agent 10 & 0.1271/0.1482 & 0.1209/0.1381 & 0.1238 (0.1151,0.1324)/0.1433 (0.1356,0.1513)\\
\hline
\end{tabular}
\end{table}

Synthetic dataset enables us also to test the Stereotype-Sharing Overlay Network (SSON) --- used when no or few past interactions are available. In the experiment, we select 10 ``inexperienced'' trustors with less than 5 interactions. When encountering a review, an inexperienced trustor requests stereotypes from other, trustworthy agents (called stereotype providers); and then combines the stereotypes (see Section \ref{sec:combiningstereotype}).


Table \ref{tab:syndataMAESSON} shows the performance of StereoTrust when SSON is applied (the values at the left of `/'). The average mean absolute errors for honest agents, dishonest agents and all agents are, respectively, 0.1277, 0.1211 and 0.1242. By comparing with the results without SSON (Tab. \ref{tab:syndataMAE}), we notice that even if the trustor does not have sufficient local knowledge, by requesting other agents' stereotypes, it is still able to reasonably estimate the trustworthiness of the potential interaction partner. In order to further validate the usefulness of the SSON, we let the 10 selected inexperienced trustors collect third party stereotypes randomly for trust estimation (i.e., SSON is not applied). On the average, SSON lowers MAE in comparison with random selection by around 13.22\%.
This result demonstrates that SSON evidently improves the performance of StereoTrust when trustor's local knowledge is insufficient.


\subsubsection{Discussion}
d-StereoTrust has the highest prediction accuracy (the smallest MAE) at the cost of incomplete coverage (95.5\%, Table~\ref{tab:syndataMAE}). Moreover, d-StereoTrust requires only fragmentary third-party information, as the trustor only asks agents that are also interested in corresponding categories. Consequently, d-StereoTrust is robust even when up to 40\% of agents are malicious.

EigenTrust not only has lower prediction accuracy (the difference, albeit small, is statistically significant), but also requires a complex, distributed calculation, thus incurring high communication overhead. Additionally, EigenTrust requires some pre-trusted agents, which may not exist in reality.

d-StereoTrust uses both stereotypes and historic information. As the accuracy of dichotomy-only is lower, stereotypes indeed improve the prediction accuracy.

Other standard methods, the feedback aggregation and both variations of transitive trust models, have significantly lower prediction accuracy. Moreover, the transitive trust model (using the most reliable path) has the lowest coverage. BLADE model improves prediction accuracy by learning rating providers' bias. However, BLADE requires that the trustor must have sufficient experience with the rating providers such that the target agent's trustworthiness can be reliably inferred. This makes BLADE ineffective in some cases, and hence lowering its overall accuracy.


Although StereoTrust has lower prediction accuracy than d-StereoTrust, StereoTrust uses only local information, and no opinions of third-parties. The key to accuracy here are the appropriate stereotypes --- the more the stereotypes mirror the true honesty of the agents, the more accurate predictions StereoTrust will form. We expect that in some contexts such stereotypes can be evaluated by, or formed with, an assistance of a human operator.

With SSON (Table~\ref{tab:syndataMAESSON}), the trustors without sufficient local knowledge can predict trust by requesting other agents' stereotypes. Thus, the coverage of StereoTrust model becomes complete. Since other agents may provide fake stereotypes maliciously, some of the collected stereotypes may not derive accurate trust. However, by updating trust scores of stereotype providers based on past accuracy of their reported stereotypes, the final aggregated stereotype information is still able to reasonably predict trustworthiness of the unknown agents.

\section{Conclusion}
\label{sec:conclusion}

We consider the problem of predicting trustworthiness of an unknown agent in a large-scale distributed setting. Traditional approaches to this problem derive unknown agent's trust essentially by combining trust of third parties to the agent with the trustor's trust of these third parties; or simply by aggregating third parties' feedbacks about the unknown agent. In contrast, StereoTrust uses different \emph{kind} of information: that of \emph{semantic} similarity of the unknown agent to other agents that the trustor personally knows. In StereoTrust,  a trustor builds stereotypes that aggregate and summarize the experience it had with different kinds of agents. The criteria by which the stereotypes are constructed are very flexible. For instance, stereotypes can be based on information from agents' personal profiles, or the class of transactions they make. So one basic assumption of StereoTrust is that such profile information is correctly available. We believe this is a reasonable assumption because it is rare to interact with an agent about which absolutely no information is
available. Facing a possible transaction with an unknown agent, the trustor estimates its trust by cumulating the experience from the stereotypes to which the unknown agent conforms.

The stereotypes are based on the local perspective and local information of the trustor, and, therefore, are naturally suited for large-scale systems; personalized for each trustor; and less susceptible to false or unsuitable information from third parties.

When some third parties' opinions about an agent are available, we propose an enhancement (d-StereoTrust), which creates a ``good'' and a ``bad'' subgroup inside each stereotype. The trustor assigns each one of its previous transaction partners to one of these groups based on its personal experience with the partner (e.g., the ratio of failed transactions). Then, the trustor uses the aggregated third parties' opinions about the unknown agent to determine how similar is the agent to the ``good'' and the ``bad'' subgroup. Third parties' opinions are a small subset of information used by traditional mechanisms (such as feedback aggregation or Eigentrust-type algorithms).
According to our experiments, by combining stereotypes with the partial historic information, d-StereoTrust predicts the agent's behavior more accurately than Eigentrust and feedback aggregation.

StereoTrust can be not only personalized for a particular trustor, but also for a particular type of interactions (classified by groups). We are currently working on such extensions of StereoTrust, as well as exploring possible concrete applications, including a P2P storage system.

While our technique is novel in the context of evaluating trust -- and provides a new paradigm of using stereotypes for trust calculation instead of using feedbacks or a web of trust -- it bears resemblance with collaborative filtering techniques \cite{cfsurvey}. The primary difference is that StereoTrust uses only local information in a decentralized system. However, the similarities also mean that while our work proposes a new paradigm to determine trust, the methodology we use is not out of the blue. Also, we anticipate that sophisticated collaborative filtering as well as machine learning techniques can be adopted to further improve StereoTrust's performance. Some nascent attempts in this direction include our recent works \cite{metatrustjournal}.

As mentioned in Section \ref{sec:discussion}, StereoTrust is not explicitly designed to handle the scenario where the agents' behavior may change over time. However, by studying relevant contextual information for stereotype formation, such a problem can be partially addressed. As part of future work, we intend to work on incorporating various dynamic approaches (such as learning agents' behavior pattern) to the model.



\end{document}